\documentclass[useAMS,usenatbib]{mn2e}

\usepackage{revsymb}
\usepackage{amsmath}
\usepackage{amsfonts}
\usepackage{amssymb}
\usepackage{graphicx}

\title[The circles-in-the-sky test]
  {A search for cosmic topology in the final WMAP data}
\author[R. Aurich and S. Lustig]
  {R.~Aurich and S.~Lustig \\
  Institut f\"ur Theoretische Physik, Universit\"at Ulm,\\
  Albert-Einstein-Allee 11,\\ D-89069 Ulm, Germany
}

\date{}

\pagerange{\pageref{firstpage}--\pageref{lastpage}} \pubyear{2012}

\def\LaTeX{L\kern-.36em\raise.3ex\hbox{a}\kern-.15em
    T\kern-.1667em\lower.7ex\hbox{E}\kern-.125emX}

\begin{document}

\def\bfis{\hbox{\scriptsize\rm i}}
\def\bfi{\hbox{\rm i}}
\def\bfj{\hbox{\rm j}}

\newcommand{\apj}{{Astrophys.\ J.}}
\newcommand{\apjs}{{Astrophys.\ J.\ Supp.}}
\newcommand{\apjl}{{Astrophys.\ J.\ Lett.}}
\newcommand{\aj}{{Astron.\ J.}}
\newcommand{\prl}{{Phys.\ Rev.\ Lett.}}
\newcommand{\prd}{{Phys.\ Rev.\ D}}
\newcommand{\mnras}{{Mon.\ Not.\ R.\ Astron.\ Soc.}}
\newcommand{\araa}{{ARA\&A }}
\newcommand{\aap}{{Astron.\ \& Astrophy.}}
\newcommand{\nat}{{Nature }}
\newcommand{\cqg}{{Class.\ Quantum Grav.}}

\setlength{\topmargin}{-1cm}

\label{firstpage}

\maketitle

\begin{abstract}
A search for matched circle pairs of similar temperature fluctuations
in the final WMAP 9yr data is carried out.
Such a signature is expected if the space of the Universe is multiply connected.
We investigate the relation between the pixel resolution of
cosmic microwave background (CMB) maps
and a Gaussian smoothing in order to lower the probability for missing
matched circle pairs.
CMB maps having the 3-torus topology are generated with the
characteristics of the WMAP satellite in order to determine
how large the smoothing should be chosen in CMB maps
disturbed by detector noise.
The V- and W-band data are analysed with respect to matched circle pairs
and a tentatively signal is found for a circle pair,
which lies, however, close to the plane of the Galaxy.
It is, however, inconclusive whether this signal is generated by chance,
is due to residual foregrounds contained in the V- and W-band maps,
or is due to a genuine topology.
\end{abstract}


%%%%%%%%%%%%%%%%%%%%%%%%%%%%%%%%%%%%%%%%%%%%%%%%%%%%%%%%%%%%%%%%%%%%%%%%%%%%
%%%%%%%%%%%%%%%%%%%%%%%%%%%%%%%%%%%%%%%%%%%%%%%%%%%%%%%%%%%%%%%%%%%%%%%%%%%%

%%%%%%%%%%%%%%%%%%%%%%%%%%%%%%%%%%%%%%%%%%%%%%%%%%%%%%
\section{Introduction}
%%%%%%%%%%%%%%%%%%%%%%%%%%%%%%%%%%%%%%%%%%%%%%%%%%%%%%

Since the cosmic microwave background (CMB) radiation provides us with
the earliest admissible electromagnetic radiation,
it contains information about the largest scales of our Universe.
Therefore, it is most promising to search for topological signatures
in the CMB radiation in order to reveal the topology of our Universe.
One of the topological tests is the matched circle test
proposed by \cite{Cornish_Spergel_Starkman_1998b}
on which this paper puts its focus.
For an introduction into the topic of cosmic topology and
discussions concerning topological tests, see
\cite{Lachieze-Rey_Luminet_1995,Luminet_Roukema_1999,Levin_2002,%
Reboucas_Gomero_2004,Luminet_2008,Mota_Reboucas_Tavakol_2010,%
Mota_Reboucas_Tavakol_2011,Fujii_Yoshii_2011}.

The idea behind this test is the following.
The CMB sky observed from a given observer position originates
from a sphere around this observer,
i.\,e.\ from the surface of last scattering (SLS).
If the Universe possesses a non-trivial topology,
the space can be viewed as being tessellated by cells
which have to be identified.
This space is the so-called universal cover.
Each cell has a ``clone'' of the observer,
which observes the same CMB sky.
If the observer and a given clone are not farther separated
than the diameter of the surface of last scattering,
the two spheres overlap in the universal cover.
The intersection of the  two spheres is a circle
seen by the observer and its clone in a different direction.
Since the observer and its clone are to be identified,
one concludes that two circles should exist on the CMB sky
with identical temperature fluctuations
seen in different directions but with the same radius.
A non-trivial topology is thus betrayed by as much pairs of circles
with the same temperature fluctuations
as there are clones of the observer not farther away than
the diameter of the surface of last scattering.

This description is simplified since it ignores several effects
which alter the CMB temperature fluctuations of the observer
and its clone in a different way,
so that the temperature fluctuations on both circles are no longer identical.
The two most important CMB contributions in this respect are the
Doppler contribution,
whose magnitude depends on the velocity projection towards the observer,
and the integrated Sachs-Wolfe (ISW) contribution,
which arises along the path from the surface of last scattering
to the observer or to the clone.
These two paths are not identified and lead to different contributions
to the total CMB signal.
These issues are discussed in more detail by
\cite{Aurich_Lustig_Steiner_2005b,Riazuelo_et_al_2006,%
Key_Cornish_Spergel_Starkman_2007,Bielewicz_Banday_2011}.
There are further degrading effects,
e.\,g.\ the finite thickness of the surface of last scattering,
but these are considered as subdominant to the Doppler and ISW contribution.
Despite these degrading effects,
there should be a detectable topological signal.

In addition to these contributions,
which can be computed with the standard CMB physics, 
there are residuals left over by the subtraction of foreground sources
which have their own uncertainties. 

The circles-in-the-sky (CITS) test requires a full sky survey and has been
applied to different sky maps derived from the COBE and WMAP missions.
\cite{Roukema_1999,Roukema_2000} applied the CITS test to the COBE data.
\cite{Cornish_Spergel_Starkman_Komatsu_2003} and
\cite{Key_Cornish_Spergel_Starkman_2007}
analyse the first year CMB data with respect to
nearly back-to-back circle pairs with a negative result.
A hint of the Poincar\' e dodecahedral topology is found by
\cite{Roukema_et_al_2004} in the first year data.
The circle signals correspond to circles with a small radius of
about $11^\circ$.
This claim could not be confirmed by \cite{Key_Cornish_Spergel_Starkman_2007}.
The analysis of \cite{Aurich_Lustig_Steiner_2005b} uses a
weight function constructed to suppress the Doppler and ISW contributions.
It does not find a convincing signal, and only a marginal hint is found
in favour of the Poincar\' e dodecahedral space.
The three-year WMAP data are investigated with respect to the 3-torus topology
by \cite{Aurich_Janzer_Lustig_Steiner_2007}, and it is found
that the degree of uncertainties in the CMB map must be
significantly below 50$\mu\hbox{K}$
in order to have a realistic chance to discover such a topology.
The seven year WMAP data are analysed by \cite{Bielewicz_Banday_2011}
with respect to back-to-back circle pairs.
They rule out topologies having such circle pairs with radii
larger than $10^\circ$.
This work emphasises the presence of residual Galactic foreground emission
close to the Galactic plane in the WMAP ILC 7yr map and
chooses to use a mask to eliminate this non-CMB contribution.
The search for general circle pairs that are not back-to-back
is carried out by \cite{Vaudrevange_Starkman_Cornish_Spergel_2012}.
Although some signals are detected, they are ascribed to the residual
foreground contained in the ILC map which is used within the WMAP Kp12 sky mask.
This analysis concludes that no topological signal is present in the
seven year WMAP data with circle radii above $10^\circ$.
A CITS search for an orbifold line topology is carried out by
\cite{Rathaus_BenDavid_Itzhaki_2013} and a tentative candidate match is found.

Are those topological spaces definitely ruled out by these CMB analyses?
In order to exclude or find possible loop holes in the matched circle pair
signal,
we generate a high resolution CMB sky map for the 3-torus topology
which requires a space with zero curvature, i.\,e.\ an Euclidean space. 
This map is analysed with respect to the amplitude of the topological signal.

The 3-torus map is computed using the cosmological parameters
of the $\Lambda$CDM concordance model
which are published by \cite{Jarosik_et_al_2010} in their Table 8,
column ``WMAP+BAO+$H_0$''
($\Omega_{\hbox{\scriptsize bar}} = 0.0456$,
$\Omega_{\hbox{\scriptsize cdm}} = 0.227$, 
Hubble constant $h=0.704$, reionization optical depth $\tau=0.087$,
scalar spectral index $n_s=0.963$).
In the following only the cubic 3-torus is considered
where all three topological lengths have the same value $L$.
In units of the Hubble length $L_{\hbox{\scriptsize H}} = c/H_0$,
the chosen side length is $L=1.5$ in sections \ref{Individual_CITS_amplitude}
to \ref{search_grid}.
This value is significantly smaller than that obtained from the
requirement that the CMB temperature correlations of the model
should match the observed correlations.
A best fit of the correlations leads to values around $L=4$,
see figure 11 in \cite{Aurich_Lustig_2010b}.
It should be noted that this 3-torus topology with $L=4$ could be
detected in future high-redshift galaxy surveys as emphasized by
\cite{Roukema_France_Kazimierczak_Buchert_2013}
which would be independent of the CMB analyses.
The reason for the unrealistically small value of $L=1.5$ is
that the CMB anisotropies have to be computed by using the eigenmodes
of the 3-torus, and we want to accurately compute the CMB
up to the multipole moment $C_l$ with $l_{\hbox{\scriptsize max}}=3000$.
To achieve this fine-structure for the chosen cosmological parameters,
all eigenmodes with a wavenumber $k$ up to $k=1\,026$ are included.
This means that 61\,556\,892 different wavenumbers $\vec k$
are taken into account.
The simulation of a CMB map for $L=4$ with that resolution is
currently not feasible.
However, the amplitude of the signal due to a single pair of matched circles
should be independent of the size of the fundamental cell.
The size determines, however, the number of matched circle pairs
which decreases with increasing size of the 3-torus cell.
The cubic torus topology with $L=4$ possesses
six circle pairs with radius $\alpha\simeq 31^\circ$ and
three  circle pairs with radius $\alpha\simeq 53^\circ$.
The total number of circle pairs increases to 182 for $L=1.5$.
Thus, the simulation with $L=1.5$ has the advantage
that there are much more circle pairs to analyse with respect to
their discovery probability.
For the decision, whether a cell with $L=4$ is discovered with certainty,
one has to take the much smaller number of circle pairs into account.
The CMB sky map for the 3-torus with $L=1.5$ is generated in the HEALPix format
\citep{Gorski_Hivon_Banday_Wandelt_Hansen_Reinecke_Bartelmann_2005}
with a resolution of $N_{\hbox{\scriptsize side}}=4096$.
This map possesses a pixel size of $51''.5$ which is three times
smaller than the resolution of the simulation
being of the order of $180^\circ/l_{\hbox{\scriptsize max}} \sim 3'.6$.
The following analysis is mainly based on this map.
We will also use less accurate $L=4$ simulations
in sections \ref{Noise_CITS_amplitude} and \ref{CITS_signal_with_masks}
when the detector noise is taken into account
since in that case a lower accuracy of the simulation is sufficient.
Furthermore, we restrict us to a search of back-to-back circle pairs.

%%%%%%%%%%%%%%%%%%%%%%%%%%%%%%%%%%%%%%%%%%%%%%%%%%%%%%
\section{The CITS amplitude for individual circle pairs}
%%%%%%%%%%%%%%%%%%%%%%%%%%%%%%%%%%%%%%%%%%%%%%%%%%%%%%
\label{Individual_CITS_amplitude}

Let us now turn to the correlation measure used for
the detection of CITS signals.
The expansion of the temperature $\delta T_i(\phi)$ along a circle in a Fourier
series $\delta T_i(\phi)=\sum_m T_{im} e^{\hbox{\scriptsize i}m\phi}$,
$0\leq \phi\leq 2\pi$,
allows one to define the $m$-weighted circle signature for
two circles $i$ and $j$ having a radius $\alpha$
as \citep{Cornish_Spergel_Starkman_Komatsu_2003}
\begin{equation}
\label{Eq:cits_m_factor}
S_{ij}(\alpha,\beta) \; := \;
\frac{2\sum_m m T_{im} T_{jm}^\star e^{\hbox{\scriptsize i}m\beta}}
{\sum_m m \big(|T_{im}|^2+|T_{jm}|^2\big)}
\hspace{10pt} .
\end{equation}
The angle $\beta$ takes a possible shift between the two circles into account.
A perfect correlation would be revealed by $S_{ij}(\alpha,\beta)=1$,
but due to the degrading Doppler and ISW contributions,
the actual value is lower.
Then the maximum is taken over all circle pairs and all shift angles $\beta$
for a fixed circle radius $\alpha$, i.\,e.\
\begin{equation}
\label{Eq:cits}
S(\alpha) \; = \; \max_{i,j,\beta}\, S_{ij}(\alpha,\beta)
\hspace{10pt} .
\end{equation}
In the following we restrict this analysis to back-to-back circles
which determines the index $j$ as a function of $i$.
Every pixel $i$ of the HEALPix map is treated as a possible
centre of a circle in the CITS search.

%%%%%%%%%%%%%%%%%%%%%%%%%%%%%%%%%%%%%%%%%%%%%%%%%%%%%%%%%%%%%%%%%%%%%%%%%%%%
%%%%%%%%%%%%%%%%%%%%%%%%%%%%%%%%%%%%%%%%%%%%%%%%%%%%%%%%%%%%%%%%%%%%%%%%%%%%

\begin{figure}
\begin{center}
\begin{minipage}{10cm}
\vspace*{-25pt}
\hspace*{-20pt}\includegraphics[width=10.0cm]{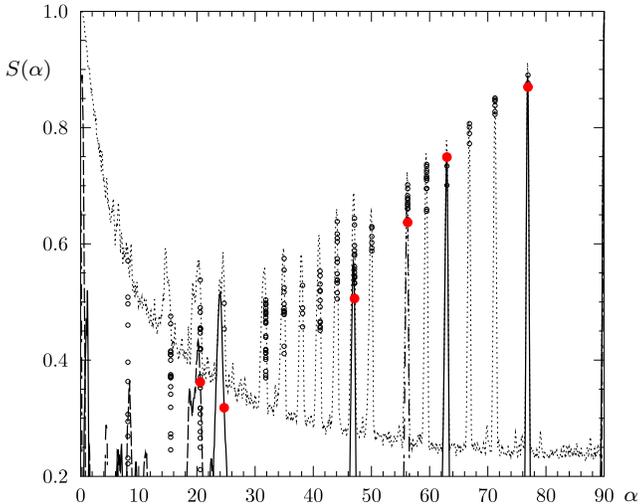}
\put(-36,29){$\alpha$}
\put(-270,190){$S(\alpha)$}
\end{minipage}
\vspace*{-25pt}
\end{center}
\caption{\label{Fig:CITS_for_single_rings}
The CITS correlation $S(\alpha)$ is calculated for a 3-torus simulation
with $L=1.5$ using the HEALPix resolution $N_{\hbox{\scriptsize side}}=512$
and a Gaussian smoothing with $\theta=0.5^\circ$.
The dotted curve shows $S(\alpha)$ obtained from all back-to-back circle pairs
and all shift angles $\beta$.
The open circles reveal the CITS values obtained from the
182 matched circle pairs of this topology.
The dashed, full and dot-dashed curves belong to the subset of
circle pairs having the same circle centres as one of the matched circle pairs
with radii $\alpha=20.5^\circ$, $24.7^\circ$, and $56.2^\circ$, respectively.
The CITS values of these matched circle pairs are depicted by red dots.
}
\end{figure}

%%%%%%%%%%%%%%%%%%%%%%%%%%%%%%%%%%%%%%%%%%%%%%%%%%%%%%%%%%%%%%%%%%%%%%%%%%%%
%%%%%%%%%%%%%%%%%%%%%%%%%%%%%%%%%%%%%%%%%%%%%%%%%%%%%%%%%%%%%%%%%%%%%%%%%%%%

Even if the correlation $S(\alpha)$ is computed for a CMB map
belonging to a model with a simply-connected topology,
one obtains non-vanishing correlations by chance.
The smaller the rings, the higher is the probability that the
temperature fluctuations are accidentally correlated.
This leads to a background for $S(\alpha)$,
which starts at $S(\alpha)\simeq 1$ for $\alpha=0$ and declines
towards larger radii $\alpha$,
until at $\alpha=90^\circ$ the pairs are mapped on to each other
leading again to $S(\alpha)\simeq 1$.
A topology can only be discerned if the CITS correlation due to
a matched circle pair is significantly larger than this background.

This background behaviour is shown in figure \ref{Fig:CITS_for_single_rings},
where the CITS correlation $S(\alpha)$ is computed from a 3-torus simulation
with $L=1.5$ as described in the Introduction.
The dotted curve belongs to $S(\alpha)$ obtained as the maximum
taken over all occurring back-to-back circle pairs and all shifts $\beta$.
Even for very large radii $\alpha$, the background is not far below 0.3.
Since there are 182 matched circle pairs, every peak in $S(\alpha)$ is
generated by the circle pair with the highest signal.
The figure also reveals the variation in the CITS signal for the
individual pairs.
For each pair the individual CITS value is depicted by a small circle
in figure \ref{Fig:CITS_for_single_rings}.
For $\alpha \lesssim 20^\circ$ the CITS values of matched circle pairs
are below  the background of non-matched circle pairs,
so that such small matched circle pairs can probably not be detected.
Above $\alpha \simeq 25^\circ$ the matched circle pairs can be detected
under the optimal conditions assumed here,
where a pure CMB map without foregrounds and detector noise is available.

Figure \ref{Fig:CITS_for_single_rings} also shows the
CITS correlation $S(\alpha)$ obtained from the subset of circle pairs
having all the same circle centres.
For such a subset, the background due to non-matched circle pairs
is much lower and not visible in figure \ref{Fig:CITS_for_single_rings}.
The CITS values of three subsets of such circle pairs are plotted
with radii $\alpha=20.5^\circ$, $24.7^\circ$, and $56.2^\circ$.
The CITS values of the matched circle pairs
(red dots in figure \ref{Fig:CITS_for_single_rings})
with $\alpha=20.5^\circ$ and $24.7^\circ$
are below the full background (dotted curve).
But in their neighbourhood towards smaller radii,
there are correlations above the background level.
This is a numerical observation which is confirmed in many instances:
The maximal values in the CITS correlation are not obtained exactly on
the matched circle pair, but, instead, for pairs shifted towards smaller radii.
This implies that if one would discover a CITS signal in the real sky map,
the radius $\alpha$ of the true circle pair is probably slightly larger
than that corresponding to the peak.
Along the axis of the selected $24.7^\circ$ matched circle pair are three
further pairs at radii $47.1^\circ$, $63.0^\circ$, and $76.9^\circ$,
see the full curve and the corresponding red dots
in figure \ref{Fig:CITS_for_single_rings}.
One observes that the shift towards smaller ring radii is less pronounced
if the ring radius $\alpha$ increases.
Although this shift is only a numerical observation,
one might speculate that the signal of a matched circle pair is accidentally
enhanced by pixels close to the pair.
Since a circle slightly inside the considered matched circle has fewer pixels,
it possesses a higher probability to corroborate the positive signal of
the matched pair than a slightly larger circle.
In addition, this effect should diminish for larger circle radii $\alpha$
as it is observed.

%%%%%%%%%%%%%%%%%%%%%%%%%%%%%%%%%%%%%%%%%%%%%%%%%%%%%%
\section{Smoothing and the CITS amplitude}
%%%%%%%%%%%%%%%%%%%%%%%%%%%%%%%%%%%%%%%%%%%%%%%%%%%%%%
\label{Smoothing_CITS_amplitude}

Let us now turn to the important question
how the smoothing of a CMB map can improve the amplitude of
a CITS signal of a matched circle pair
so that its peak emerges out of the background.
This can be essential for detecting a possible topology.
There are several reasons why a smoothing of the map should be carried out.
On the one hand, the temperature values of the pixels contain not
only the pure CMB signal, but in addition the noise of the detector.
Assuming that the noise is independent from pixel to pixel,
a smoothing suppresses the noise contribution leading to a cleaner
CMB signal.
Even without noise a smoothing can be advantageous.
Since the data are given on a discretised map,
the scan strategy can miss the very localised peaks
if the search grid is too coarse.
If the data are sufficiently smoothed,
even a relatively coarse search grid can find the topological signal,
and a lot of computer time can be saved.
However, a too strong smoothing will eliminate the CITS signal.
Thus there is a relation of how strong the smoothing should be
for a given HEALPix resolution $N_{\hbox{\scriptsize side}}$.
As discussed in the Introduction, the CMB signal is composed of
different contributions and only the Sachs-Wolfe contribution
leads to a clear CITS signal
while the Doppler and integrated Sachs-Wolfe contributions deteriorate it.
The Sachs-Wolfe contribution provides the largest signal on scales
around the first acoustic peak having a scale of roughly $0.8^\circ$.
Thus the smoothing should at least be smaller than that in order to
preserve the information due to the Sachs-Wolfe contribution.

%%%%%%%%%%%%%%%%%%%%%%%%%%%%%%%%%%%%%%%%%%%%%%%%%%%%%%%%%%%%%%%%%%%%%%%%%%%%
%%%%%%%%%%%%%%%%%%%%%%%%%%%%%%%%%%%%%%%%%%%%%%%%%%%%%%%%%%%%%%%%%%%%%%%%%%%%

\begin{figure}
\begin{center}
\begin{minipage}{10cm}
\vspace*{-25pt}
\hspace*{-20pt}\includegraphics[width=10.0cm]{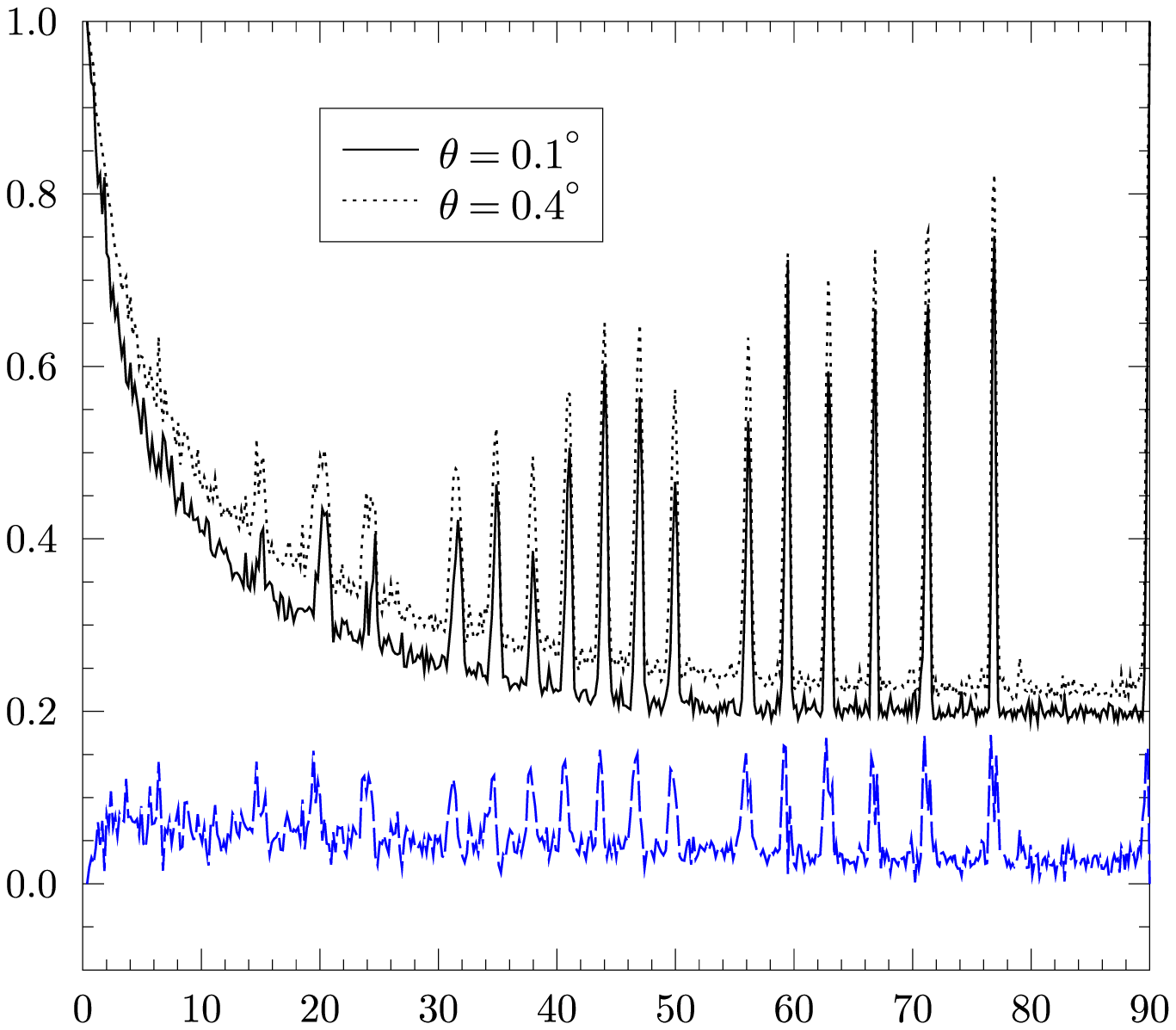}
\put(-36,29){$\alpha$}
\put(-270,195){$S(\alpha)$}
\put(-225,200){(a)}
\put(-100,200){$N_{\hbox{\scriptsize side}}=256$}
\end{minipage}
\begin{minipage}{10cm}
\vspace*{-50pt}
\hspace*{-20pt}\includegraphics[width=10.0cm]{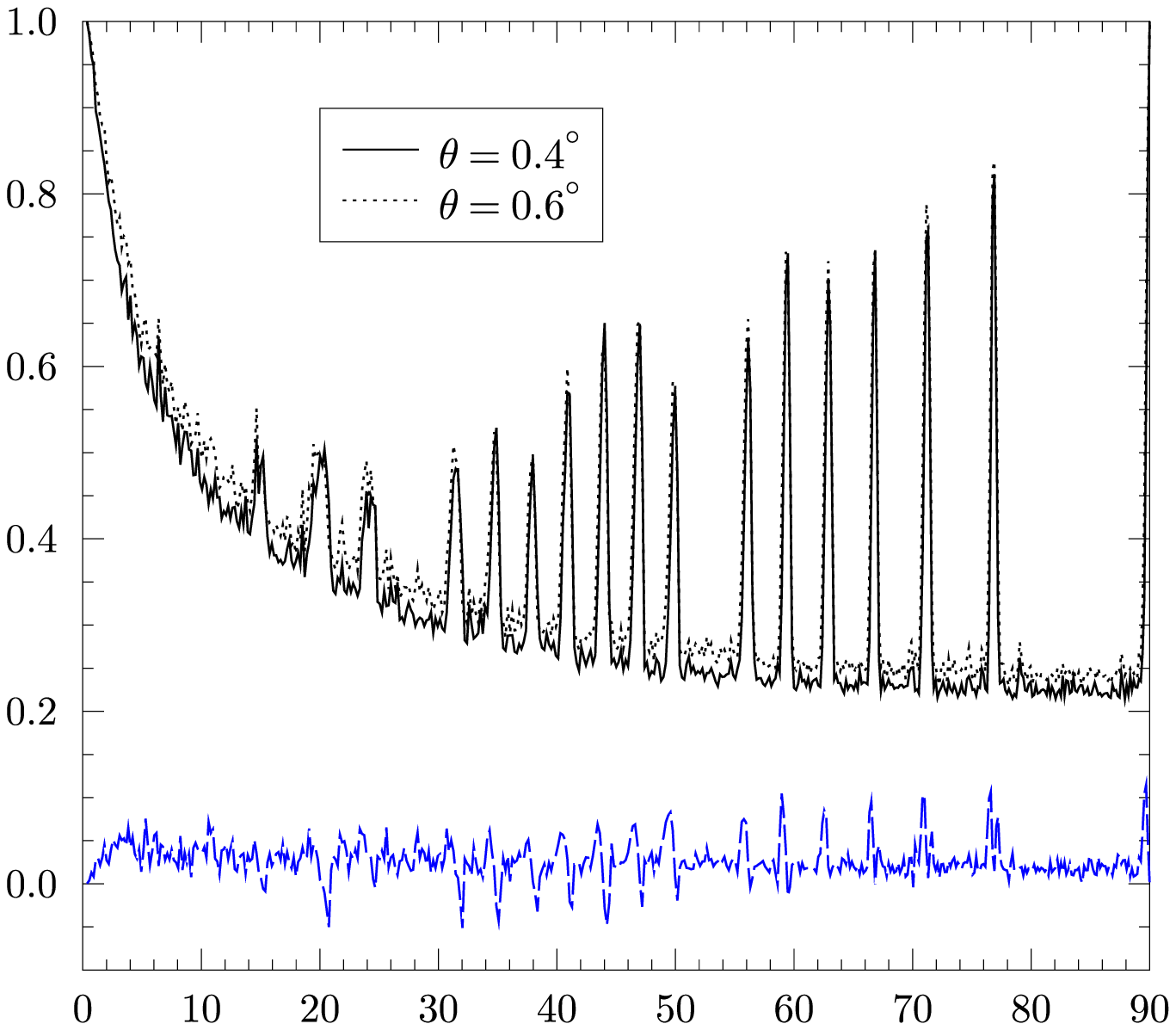}
\put(-36,29){$\alpha$}
\put(-270,195){$S(\alpha)$}
\put(-225,200){(b)}
\put(-100,200){$N_{\hbox{\scriptsize side}}=256$}
\end{minipage}
\vspace*{-25pt}
\end{center}
\caption{\label{Fig:CITS_Gaussian_Smoothing_256}
The CITS correlation $S(\alpha)$ is calculated for a 3-torus simulation
with $L=1.5$ using the HEALPix resolution $N_{\hbox{\scriptsize side}}=256$
and three different Gaussian smoothing.
Panel (a) compares the CITS correlation for the two smoothing parameters
$\theta=0.1^\circ$ and $\theta=0.4^\circ$,
while panel (b) compares $\theta=0.4^\circ$ and $\theta=0.6^\circ$.
The dashed curve, which starts for $\alpha=0$ at zero,
displays the difference between the two CITS correlations $S(\alpha)$
shown in the panel.
}
\end{figure}

%%%%%%%%%%%%%%%%%%%%%%%%%%%%%%%%%%%%%%%%%%%%%%%%%%%%%%%%%%%%%%%%%%%%%%%%%%%%
%%%%%%%%%%%%%%%%%%%%%%%%%%%%%%%%%%%%%%%%%%%%%%%%%%%%%%%%%%%%%%%%%%%%%%%%%%%%

We use a Gaussian smoothing which multiplies the expansion coefficients
$a_{lm}$ of the temperature fluctuations $\delta T(\hat n)$ with
respect to spherical harmonics $Y_{lm}(\hat n)$ by
\begin{equation}
\label{Eq:Gaussian}
a_{lm}^{\hbox{\scriptsize sm}} \; = \;
a_{lm} \, e^{-\frac{l(l+1)\sigma^2}2}
\hspace{5pt} \hbox{ with } \hspace{5pt}
\sigma = \frac{\pi\,\theta/180^\circ}{2\sqrt{2\log 2}}
\hspace{10pt} ,
\end{equation}
where $\theta$ is the smoothing in degrees.

This Gaussian smoothing is applied to the 3-torus simulation with $L=1.5$.
The simulated CMB map is smoothed using the map in the HEALPix resolution
$N_{\hbox{\scriptsize side}}=4096$.
After the Gaussian smoothing, the map is downgraded to
$N_{\hbox{\scriptsize side}}=256$, and the CITS correlation $S(\alpha)$
is calculated using this $N_{\hbox{\scriptsize side}}=256$ map.
The result is shown in figure \ref{Fig:CITS_Gaussian_Smoothing_256},
where the smoothing parameters $\theta=0.1^\circ$, $\theta=0.4^\circ$,
and $\theta=0.6^\circ$ are used.
The panel (a) displays the correlation obtained
from the $\theta=0.1^\circ$ and $\theta=0.4^\circ$ maps.
The background of $S(\alpha)$ increases by increasing the smoothing parameter
$\theta$, but on the other hand, also the CITS signal of the matched
circle pair increases.
In order to decide, whether there is an improvement
by increasing $\theta$ from $\theta=0.1^\circ$ to $\theta=0.4^\circ$,
the figure also shows the difference
$S_{\theta=0.4^\circ}(\alpha) - S_{\theta=0.1^\circ}(\alpha)$ as a dashed curve.
The difference curve reveals the increasing background but also
that the peaks are more pronounced for $\theta=0.4^\circ$,
since the peaks are higher than the background.
Thus the height of the CITS peaks increases more than the background.
Increasing the smoothing further to $\theta=0.6^\circ$ does not
lead to an improvement as shown in panel (b).
In panel (b) the difference curve
$S_{\theta=0.6^\circ}(\alpha) - S_{\theta=0.4^\circ}(\alpha)$
shows for ring radii $\alpha \lesssim 50^\circ$ dips towards negative values.
Thus the increase of the background is larger than the gain in the
peak heights.
Even for very large rings there is no genuine gain in the height of the peaks,
but, instead, they are only extended towards smaller values of $\alpha$.
The focus of the smoothing should be put on the improvement of the signal
for medium rings.
Therefore, the smoothing parameter $\theta$ should be smaller than $0.6^\circ$.
We have investigated other smoothing parameters $\theta$ in this way
and conclude that for $N_{\hbox{\scriptsize side}}=256$,
the best smoothing parameter is about $\theta=0.4^\circ$.

%%%%%%%%%%%%%%%%%%%%%%%%%%%%%%%%%%%%%%%%%%%%%%%%%%%%%%%%%%%%%%%%%%%%%%%%%%%%
%%%%%%%%%%%%%%%%%%%%%%%%%%%%%%%%%%%%%%%%%%%%%%%%%%%%%%%%%%%%%%%%%%%%%%%%%%%%

\begin{figure}
\begin{center}
\begin{minipage}{10cm}
\vspace*{-25pt}
\hspace*{-20pt}\includegraphics[width=10.0cm]{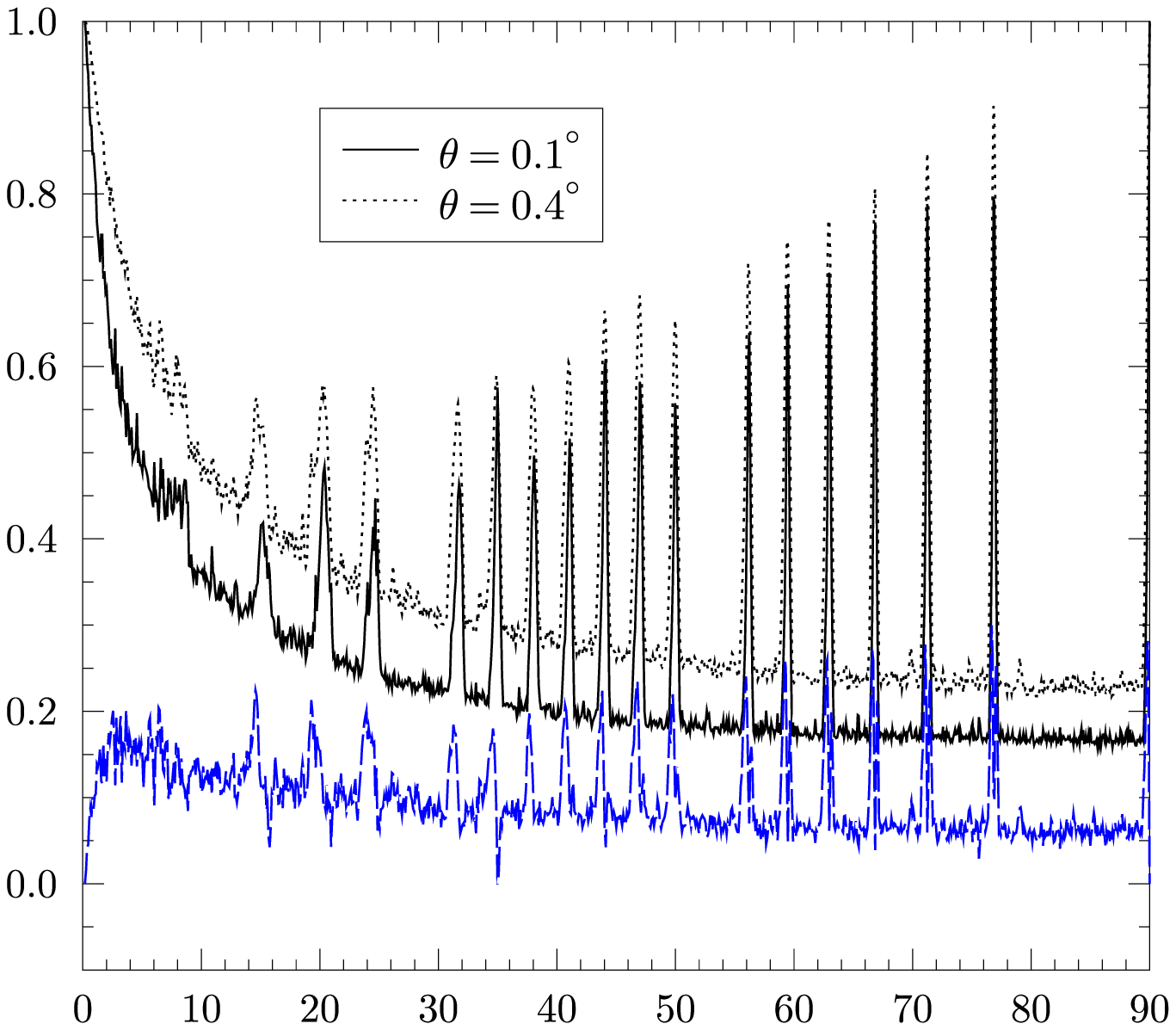}
\put(-36,29){$\alpha$}
\put(-270,195){$S(\alpha)$}
\put(-225,200){(a)}
\put(-100,202){$N_{\hbox{\scriptsize side}}=512$}
\end{minipage}
\begin{minipage}{10cm}
\vspace*{-50pt}
\hspace*{-20pt}\includegraphics[width=10.0cm]{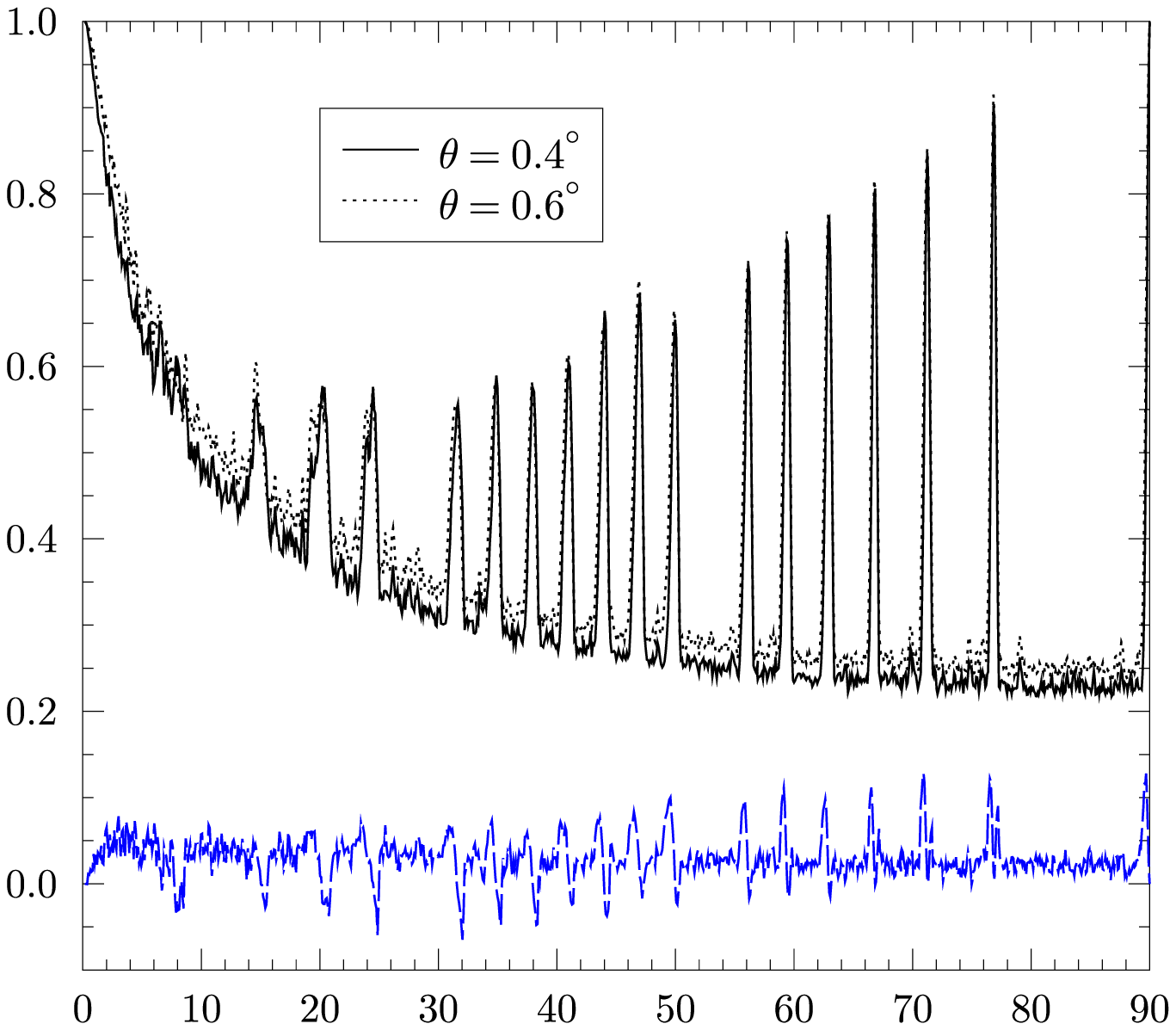}
\put(-36,29){$\alpha$}
\put(-270,195){$S(\alpha)$}
\put(-225,200){(b)}
\put(-100,202){$N_{\hbox{\scriptsize side}}=512$}
\end{minipage}
\vspace*{-25pt}
\end{center}
\caption{\label{Fig:CITS_Gaussian_Smoothing_512}
The CITS correlation $S(\alpha)$ is shown for the same simulation as
in figure \ref{Fig:CITS_Gaussian_Smoothing_256}
but now calculated from the map in the HEALPix resolution
$N_{\hbox{\scriptsize side}}=512$.
Panel (a) compares the CITS correlation for the two smoothing parameters
$\theta=0.1^\circ$ and $\theta=0.4^\circ$,
while panel (b) compares $\theta=0.4^\circ$ and $\theta=0.6^\circ$.
The dashed curve, which starts for $\alpha=0$ at zero,
displays the difference between the two CITS correlations $S(\alpha)$
shown in the panel.
}
\end{figure}

%%%%%%%%%%%%%%%%%%%%%%%%%%%%%%%%%%%%%%%%%%%%%%%%%%%%%%%%%%%%%%%%%%%%%%%%%%%%
%%%%%%%%%%%%%%%%%%%%%%%%%%%%%%%%%%%%%%%%%%%%%%%%%%%%%%%%%%%%%%%%%%%%%%%%%%%%

Let us now address the best choice of the smoothing parameter $\theta$
for the HEALPix resolution $N_{\hbox{\scriptsize side}}=512$.
To that aim, the simulated CMB map is again smoothed using the map in the
HEALPix resolution $N_{\hbox{\scriptsize side}}=4096$.
But after the Gaussian smoothing, the map is only downgraded to
$N_{\hbox{\scriptsize side}}=512$.
The CITS correlation $S(\alpha)$ is calculated from this
$N_{\hbox{\scriptsize side}}=512$ map and shown in figure
\ref{Fig:CITS_Gaussian_Smoothing_512} in the same way
as figure \ref{Fig:CITS_Gaussian_Smoothing_256} displays
the $N_{\hbox{\scriptsize side}}=256$ results.
Panel (a) shows the increase in the background by increasing
the smoothing from $\theta=0.1^\circ$ to $\theta=0.4^\circ$,
but at least for $\alpha\gtrsim 40^\circ$ the difference curve
reveals an improved signal.
However, for smaller ring radii the result is inconclusive
since positive as well as negative deviations from the background
are observed.
This suggests a twofold search if one insists on the
$N_{\hbox{\scriptsize side}}=512$ resolution:
choosing $\theta=0.4^\circ$ for larger rings and a correspondingly
smaller value of $\theta$ for smaller rings.
Panel (b) shows that a further increase in $\theta$ towards
$\theta=0.6^\circ$ does not improve the signal as it was already found
in the $N_{\hbox{\scriptsize side}}=256$ case.

As a concluding remark of this section,
it should be stated that a resolution of $N_{\hbox{\scriptsize side}}=128$
should only be used to find matched circle pairs with large radii
above $\alpha\gtrsim 40^\circ$.
A CITS search should thus be carried out in the
$N_{\hbox{\scriptsize side}}=256$ resolution using a
Gaussian smoothing parameter $\theta=0.4^\circ$.
If the sky map possesses no appreciable noise
such that a downgrade to $N_{\hbox{\scriptsize side}}=256$ is not necessary
as it might be the case for the Planck CMB map,
the $N_{\hbox{\scriptsize side}}=512$ resolution would be preferable.

%%%%%%%%%%%%%%%%%%%%%%%%%%%%%%%%%%%%%%%%%%%%%%%%%%%%%%
\section{The search grid}
%%%%%%%%%%%%%%%%%%%%%%%%%%%%%%%%%%%%%%%%%%%%%%%%%%%%%%
\label{search_grid}

In the CITS search every pixel of the HEALPix map is considered
as a possible circle centre of a matched circle pair.
The $N_{\hbox{\scriptsize side}}$ resolution leads to
$12 N_{\hbox{\scriptsize side}}^2$ pixels and determines
how thoroughly the circles are scanned.
However, it is not necessary to use for the circle centres,
for which the CITS amplitude is to be computed,
the same resolution as for the CMB map.
\cite{Vaudrevange_Starkman_Cornish_Spergel_2012} carry out their
CITS search on a CMB map having the resolution $N_{\hbox{\scriptsize side}}=512$,
but for the circle centres they use the coordinates of the pixel centres of
the HEALPix map with the resolution $N_{\hbox{\scriptsize grid}}=128$.
This procedure saves a lot of computer time since the number
of circle centres is reduced by a factor of 16.
Except in this section, we always use the same resolution for
the search grid and the CMB map.

%%%%%%%%%%%%%%%%%%%%%%%%%%%%%%%%%%%%%%%%%%%%%%%%%%%%%%%%%%%%%%%%%%%%%%%%%%%%
%%%%%%%%%%%%%%%%%%%%%%%%%%%%%%%%%%%%%%%%%%%%%%%%%%%%%%%%%%%%%%%%%%%%%%%%%%%%

\begin{figure}
\begin{center}
\begin{minipage}{10cm}
\vspace*{-25pt}
\hspace*{-20pt}\includegraphics[width=10.0cm]{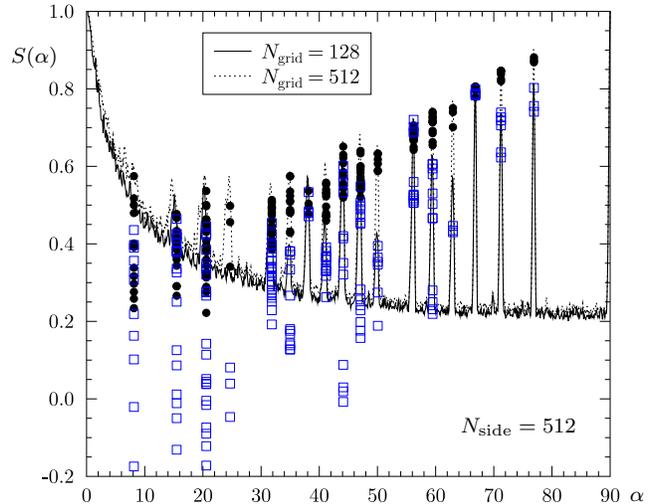}
\put(-36,29){$\alpha$}
\put(-270,195){$S(\alpha)$}
\put(-100,55){$N_{\hbox{\scriptsize side}}=512$}
\end{minipage}
\vspace*{-25pt}
\end{center}
\caption{\label{Fig:CITS_Search_Grid}
The CITS correlation $S(\alpha)$ is calculated for the 3-torus simulation
with $L=1.5$ using the HEALPix resolution $N_{\hbox{\scriptsize side}}=512$
and a Gaussian smoothing $\theta=0.4^\circ$.
The difference in the two curves arises from the resolution
$N_{\hbox{\scriptsize grid}}$ of the search grid
which provides the coordinates of the circle centres.
The dashed curve uses $N_{\hbox{\scriptsize grid}}=512$ and
the full curve $N_{\hbox{\scriptsize grid}}=128$.
Note, that both curves are computed from the same $N_{\hbox{\scriptsize side}}=512$
CMB map.
The values for the individual circle pairs are plotted as
small discs ($N_{\hbox{\scriptsize grid}}=512$) and open squares
($N_{\hbox{\scriptsize grid}}=128$).
}
\end{figure}

%%%%%%%%%%%%%%%%%%%%%%%%%%%%%%%%%%%%%%%%%%%%%%%%%%%%%%%%%%%%%%%%%%%%%%%%%%%%
%%%%%%%%%%%%%%%%%%%%%%%%%%%%%%%%%%%%%%%%%%%%%%%%%%%%%%%%%%%%%%%%%%%%%%%%%%%%

The coarser search grid harbours the risk that some matched circle pairs
get missed.
In order to test this issue, we calculate the CITS correlation $S(\alpha)$
from our 3-torus simulation with the HEALPix resolution
$N_{\hbox{\scriptsize side}}=512$ and a Gaussian smoothing $\theta=0.4^\circ$.
In the first case we use $N_{\hbox{\scriptsize grid}}=512$
and search all pixel centres of the $N_{\hbox{\scriptsize side}}=512$ map,
and in the second case only those of the $N_{\hbox{\scriptsize grid}}=128$ mesh.
The result can be inferred from figure \ref{Fig:CITS_Search_Grid}
where the dotted curve shows the $N_{\hbox{\scriptsize grid}}=512$ search
and the full curve uses the sparse $N_{\hbox{\scriptsize grid}}=128$ grid.
Several smaller rings are found only on the $N_{\hbox{\scriptsize grid}}=512$ grid,
see $\alpha=24.7^\circ$ in figure \ref{Fig:CITS_Search_Grid}.
The CITS correlations for the rings with radii $\alpha=31.8^\circ$
and $\alpha=50.0^\circ$ are reduced by a factor of two.
Taking into account that our map contains the clean CMB signal
without noise and residual foregrounds, it seems that also these
signatures can be overlooked.
Furthermore, if the multiply connected space has significantly fewer
matched circle pairs as it is the case for the $L=4$ torus topology
(six at $\alpha\simeq 31^\circ$ and three at $\alpha\simeq 53^\circ$)
mentioned in the Introduction,
even this back-to-back signature can be missed.
To emphasize the large number of matched circle pairs in the $L=1.5$ case,
the figure \ref{Fig:CITS_Search_Grid} also displays the CITS values at
the pixel centres that provide the best match to the circle centres
as small discs ($N_{\hbox{\scriptsize grid}}=512$) and open squares
($N_{\hbox{\scriptsize grid}}=128$).
It is obvious that a lot of matched circle pairs for the
$N_{\hbox{\scriptsize grid}}=128$ search are below the background
even for $\alpha\lesssim 50^\circ$.
All the open squares below the background level would never be considered
as candidates for a possible refined search.
Furthermore, since the CITS search of
\cite{Vaudrevange_Starkman_Cornish_Spergel_2012} refers to
non back-to-back circle pairs, their background level is higher
than in our back-to-back search.
This worsens the situation, so that we doubt
that their analysis excludes such matched circle pairs.

%%%%%%%%%%%%%%%%%%%%%%%%%%%%%%%%%%%%%%%%%%%%%%%%%%%%%%%%%%%%%%%%%%%%%%%%%%%%
%%%%%%%%%%%%%%%%%%%%%%%%%%%%%%%%%%%%%%%%%%%%%%%%%%%%%%%%%%%%%%%%%%%%%%%%%%%%

\begin{figure}
\begin{center}
\begin{minipage}{10cm}
\vspace*{-25pt}
\hspace*{-20pt}\includegraphics[width=10.0cm]{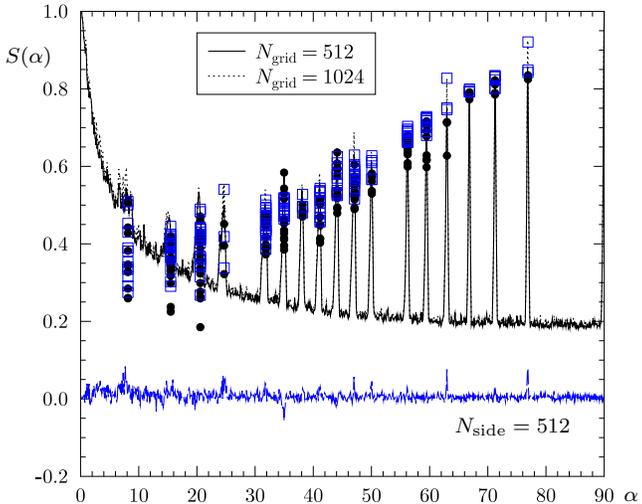}
\put(-36,29){$\alpha$}
\put(-270,195){$S(\alpha)$}
\put(-100,55){$N_{\hbox{\scriptsize side}}=512$}
\end{minipage}
\vspace*{-25pt}
\end{center}
\caption{\label{Fig:CITS_Search_Grid_Oversampling}
The CITS correlation $S(\alpha)$ is calculated for the 3-torus simulation
with $L=1.5$ using the HEALPix resolution $N_{\hbox{\scriptsize side}}=512$
and a Gaussian smoothing $\theta=0.2^\circ$.
The figure shows how the CITS signal improves if a search grid with
$N_{\hbox{\scriptsize grid}}=1024$ is used on an $N_{\hbox{\scriptsize side}}=512$
CMB map.
The results for both search grids are shown
(full curve for $N_{\hbox{\scriptsize grid}}=512$ and
dotted curve for $N_{\hbox{\scriptsize grid}}=1024$).
The values for the individual circle pairs are plotted as
small discs ($N_{\hbox{\scriptsize grid}}=512$) and open squares
($N_{\hbox{\scriptsize grid}}=1024$).
The difference curve is also shown, which fluctuates around zero.
}
\end{figure}

%%%%%%%%%%%%%%%%%%%%%%%%%%%%%%%%%%%%%%%%%%%%%%%%%%%%%%%%%%%%%%%%%%%%%%%%%%%%
%%%%%%%%%%%%%%%%%%%%%%%%%%%%%%%%%%%%%%%%%%%%%%%%%%%%%%%%%%%%%%%%%%%%%%%%%%%%

Although the computer time would rise dramatically,
one could ask how a search grid,
which is finer than the resolution of the CMB map,
would improve the CITS signature.
To address this point, the CITS correlation $S(\alpha)$ is computed
for two search grids with $N_{\hbox{\scriptsize grid}}=512$ and
$N_{\hbox{\scriptsize grid}}=1024$ on the 3-torus simulation with
the HEALPix resolution $N_{\hbox{\scriptsize side}}=512$
and a Gaussian smoothing $\theta=0.2^\circ$.
As figure \ref{Fig:CITS_Search_Grid_Oversampling} reveals
the improvement justifies the increase in computer time only for
small rings with radii $\alpha = 20^\circ\dots 30^\circ$,
where the stronger CITS signal is crucial.
A similar comparison is carried out for an $N_{\hbox{\scriptsize side}}=256$
CMB map, where two search grids with $N_{\hbox{\scriptsize grid}}=256$
and $N_{\hbox{\scriptsize grid}}=512$ are used.
It turns out that only for smoothing parameters $\theta$
which are so small that the pixelization determines the resolution
of the CMB map, an oversampling with
$N_{\hbox{\scriptsize grid}}>N_{\hbox{\scriptsize side}}$
leads to significantly better results especially for ring radii in the range
$\alpha = 20^\circ\dots 30^\circ$.

%%%%%%%%%%%%%%%%%%%%%%%%%%%%%%%%%%%%%%%%%%%%%%%%%%%%%%
\section{Noise and the CITS amplitude}
%%%%%%%%%%%%%%%%%%%%%%%%%%%%%%%%%%%%%%%%%%%%%%%%%%%%%%
\label{Noise_CITS_amplitude}

In the previous sections the CITS signature is analysed only
for CMB map simulations which contain the pure CMB signal.
Only deteriorating effects that have their origin in the CMB physics
such as the Doppler and the ISW effect were taken into account.
The experiments, in contrast, yield CMB maps which are perturbed by
noise and the beam profile of the detector.

%%%%%%%%%%%%%%%%%%%%%%%%%%%%%%%%%%%%%%%%%%%%%%%%%%%%%%%%%%%%%%%%%%%%%%%%%%%%
%%%%%%%%%%%%%%%%%%%%%%%%%%%%%%%%%%%%%%%%%%%%%%%%%%%%%%%%%%%%%%%%%%%%%%%%%%%%

\begin{figure}
\begin{center}
\begin{minipage}{10cm}
\vspace*{-25pt}
\hspace*{-20pt}\includegraphics[width=10.0cm]{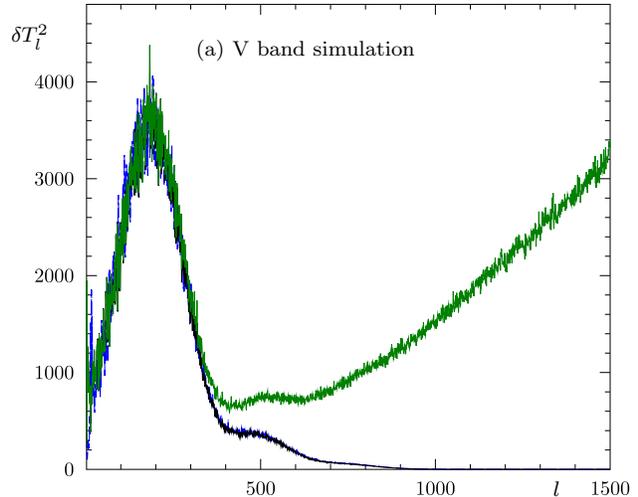}
\put(-65,28){$l$}
\put(-270,200){$\delta T_l^2$}
\put(-200,195){(a) V band simulation}
\end{minipage}
\begin{minipage}{10cm}
\vspace*{-40pt}
\hspace*{-20pt}\includegraphics[width=10.0cm]{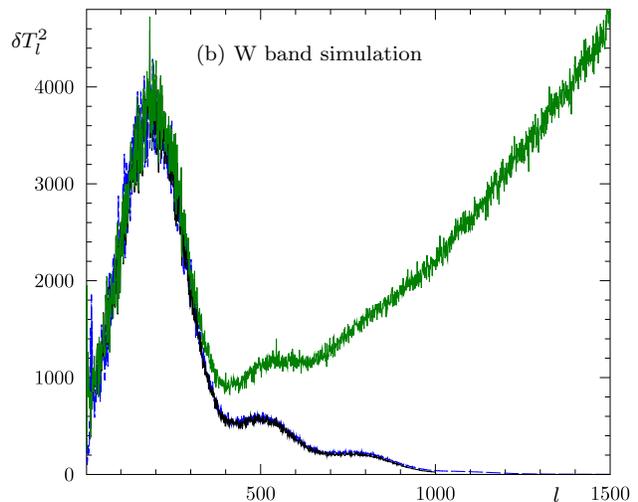}
\put(-65,28){$l$}
\put(-270,200){$\delta T_l^2$}
\put(-200,195){(b) W band simulation}
\end{minipage}
\vspace*{-25pt}
\end{center}
\caption{\label{Fig:Powerspectrum_V_and_W}
The angular power spectrum $\delta T_l^2 = l(l+1) C_l/(2\pi)$
is shown for two 3-torus simulations with $L=1.5$ and $L=4.0$.
Two power spectra belonging to $L=1.5$ and $L=4.0$ take only the beam profile
into account.
These two curves are nearly indistinguishable and drop towards zero
for large values of $l$.
A marginal difference is visible in the W band simulation for $l>1000$,
where only the $L=1.5$ simulation has power.
Furthermore, the curve increasing for large $l$ belongs to
the $L=4.0$ simulation where in addition to the beam profile
also the detector noise is accounted for.
}
\end{figure}

%%%%%%%%%%%%%%%%%%%%%%%%%%%%%%%%%%%%%%%%%%%%%%%%%%%%%%%%%%%%%%%%%%%%%%%%%%%%
%%%%%%%%%%%%%%%%%%%%%%%%%%%%%%%%%%%%%%%%%%%%%%%%%%%%%%%%%%%%%%%%%%%%%%%%%%%%

In order to address that issue,
the 3-torus CMB map is modified according to the beam profile
$b_{V1,W1}(l)$ of the W1 and V1 channels \citep{Bennett_et_al_2012}
as given on the LAMBDA website
using the HEALPix resolution $N_{\hbox{\scriptsize side}}=4096$.
Thereafter, a downgrade from $N_{\hbox{\scriptsize side}}=4096$ to
$N_{\hbox{\scriptsize side}}=512$ is carried out.
In the next step the noise of the pixel $i$ is computed from the number
$N_{\hbox{\scriptsize obs}}(i)$ of observations of the pixel $i$
as a Gaussian random error with standard deviation
$\sigma_0/\sqrt{N_{\hbox{\scriptsize obs}}(i)}$
and added to the map in the $N_{\hbox{\scriptsize side}}=512$ resolution.
This map is expanded with respect to $a_{lm}$ and the
Gaussian smoothing is realized as
\begin{equation}
\label{Eq:Gaussian_beam}
a_{lm}^{\hbox{\scriptsize sm}} \; = \;
a_{lm} \, e^{-\frac{l(l+1)\sigma^2}2} \, / \, b_{V1,W1}(l)
\hspace{10pt} ,
\end{equation}
so that the final smoothing is independent of the non-Gaussian beam profile
$b(l)$.
This sequence of operations leads to CMB maps for the 3-torus topology
comparable to the quality of the corresponding WMAP 9yr observations
in the V and W bands.

%%%%%%%%%%%%%%%%%%%%%%%%%%%%%%%%%%%%%%%%%%%%%%%%%%%%%%%%%%%%%%%%%%%%%%%%%%%%
%%%%%%%%%%%%%%%%%%%%%%%%%%%%%%%%%%%%%%%%%%%%%%%%%%%%%%%%%%%%%%%%%%%%%%%%%%%%

\begin{figure}
\begin{center}
\begin{minipage}{10cm}
\vspace*{-25pt}
\hspace*{-20pt}\includegraphics[width=10.0cm]{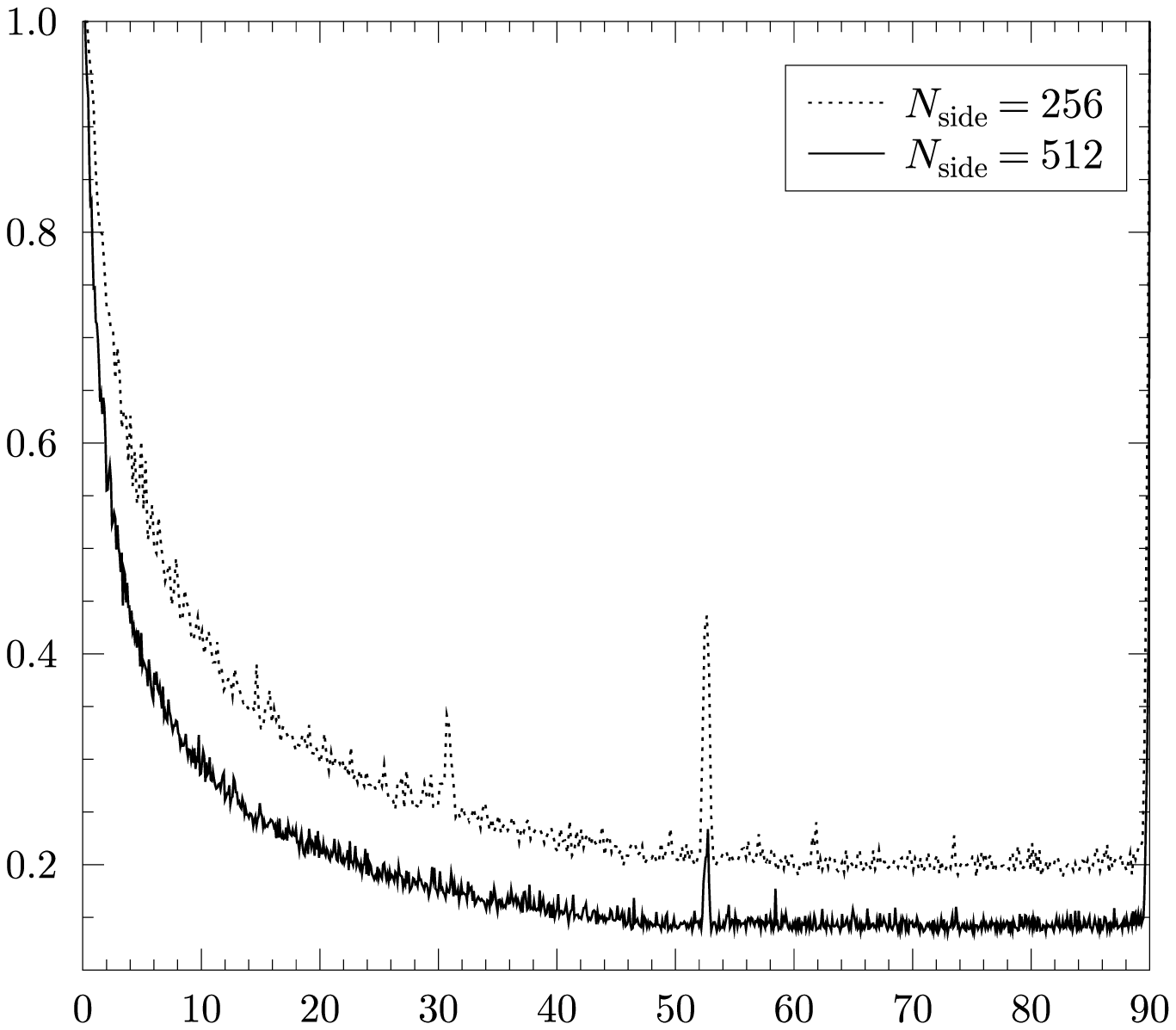}
\put(-36,29){$\alpha$}
\put(-270,195){$S(\alpha)$}
\put(-225,195){(a) V band simulation}
\end{minipage}
\begin{minipage}{10cm}
\vspace*{-40pt}
\hspace*{-20pt}\includegraphics[width=10.0cm]{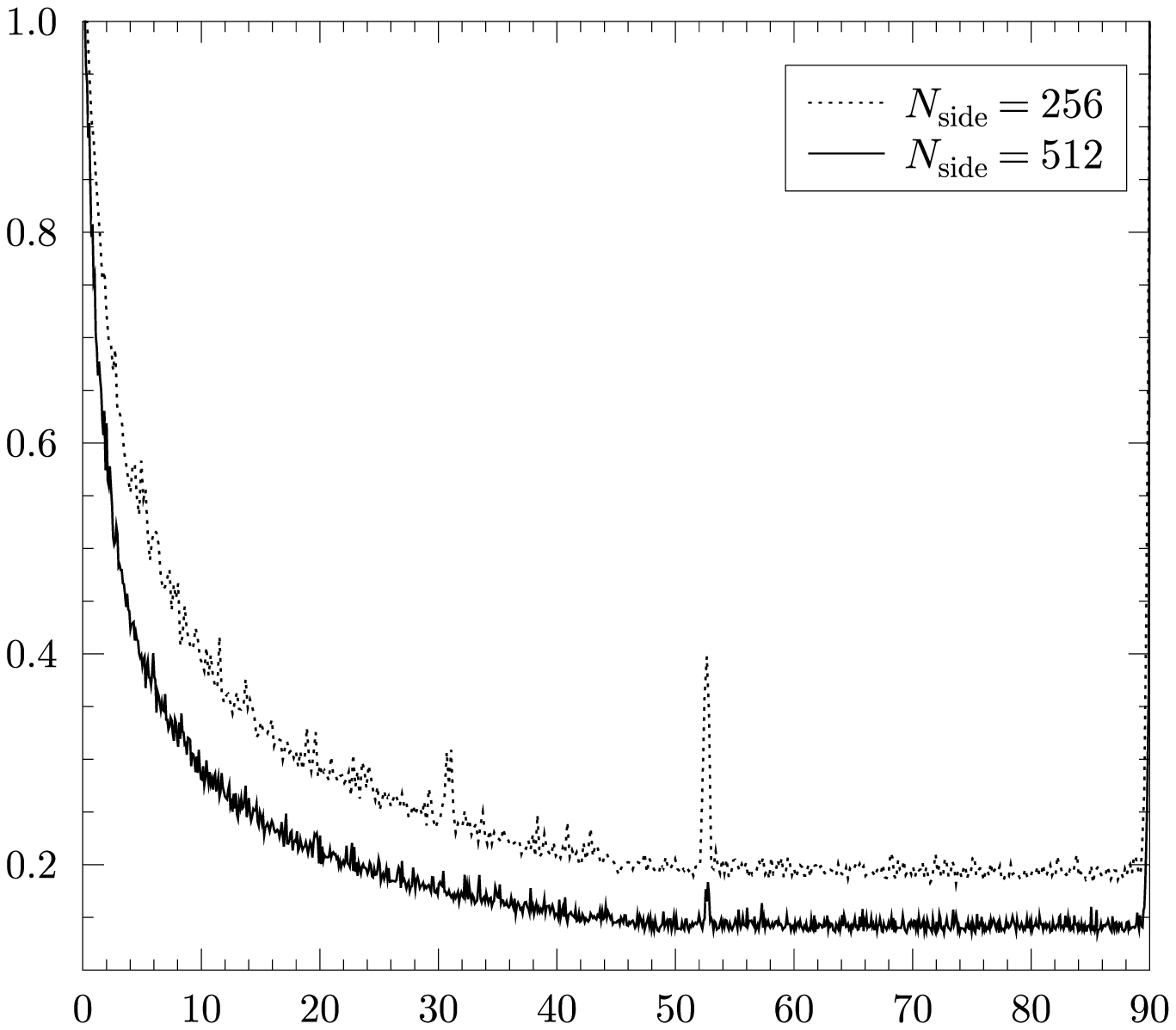}
\put(-36,29){$\alpha$}
\put(-270,195){$S(\alpha)$}
\put(-225,195){(b) W band simulation}
\end{minipage}
\vspace*{-25pt}
\end{center}
\caption{\label{Fig:CITS_V_W_Torus_Simulation}
The CITS correlation $S(\alpha)$ is calculated for the 3-torus simulation
with $L=4.0$ using the beam profile and the detector noise of
the V band in panel (a) and W band in panel (b).
The $N_{\hbox{\scriptsize side}}=256$ map is obtained by a downgrade of
the $N_{\hbox{\scriptsize side}}=512$ map.
This topology has 6 ring pairs at $\alpha=31.1^\circ$ and 3 ring pairs
at $\alpha=52.7^\circ$.
}
\end{figure}

%%%%%%%%%%%%%%%%%%%%%%%%%%%%%%%%%%%%%%%%%%%%%%%%%%%%%%%%%%%%%%%%%%%%%%%%%%%%
%%%%%%%%%%%%%%%%%%%%%%%%%%%%%%%%%%%%%%%%%%%%%%%%%%%%%%%%%%%%%%%%%%%%%%%%%%%%

In this section CMB maps for the 3-torus with side length $L=4.0$
are analysed.
In contrast to the $L=1.5$ simulation, only multipoles $l$
with $l\leq 1000$ are included in the $L=4.0$ case.
This simulation is accurate enough for an analysis with respect to noise
as can be inferred from figure \ref{Fig:Powerspectrum_V_and_W}.
For $l>1000$ the angular power spectrum $\delta T_l^2$ is noise dominated
for the V and W bands of the WMAP measurements
which are put into focus in this section.
The comparison of the two panels of figure \ref{Fig:Powerspectrum_V_and_W}
shows that the W band has a higher noise contribution than the
V band, although the W-band beam profile reveals tentatively the third
acoustic peak by ignoring the noise.

It is interesting to evaluate the CMB maps,
which are processed to mimic the V and W band maps of WMAP,
in the $N_{\hbox{\scriptsize side}}=512$ resolution without further smoothing.
Figure \ref{Fig:CITS_V_W_Torus_Simulation} shows that the CITS signature
of the 6 ring pairs at $\alpha=31.1^\circ$ is missing at the
$N_{\hbox{\scriptsize side}}=512$ resolution in both bands.
Only the 3 ring pairs at $\alpha=52.7^\circ$ emerge slightly from the
background, where the signature is more pronounced in the V band
having less noise.
The CITS signature is swamped by the noise as a simple downgrade to the
$N_{\hbox{\scriptsize side}}=256$ resolution reveals.
Here the average over 4 pixels yields the temperature value
in the $N_{\hbox{\scriptsize side}}=256$ resolution.
The improvement can be seen by the dotted curves in
figure \ref{Fig:CITS_V_W_Torus_Simulation}.
Despite the increased background, the CITS signature is visible
more clearly.
This demonstrates that the V- and W-band CMB maps have to be smoothed
in order to reduce the degrading effects due to noise.

%%%%%%%%%%%%%%%%%%%%%%%%%%%%%%%%%%%%%%%%%%%%%%%%%%%%%%%%%%%%%%%%%%%%%%%%%%%%
%%%%%%%%%%%%%%%%%%%%%%%%%%%%%%%%%%%%%%%%%%%%%%%%%%%%%%%%%%%%%%%%%%%%%%%%%%%%

\begin{figure}
\begin{center}
\begin{minipage}{10cm}
\vspace*{-25pt}
\hspace*{-20pt}\includegraphics[width=10.0cm]{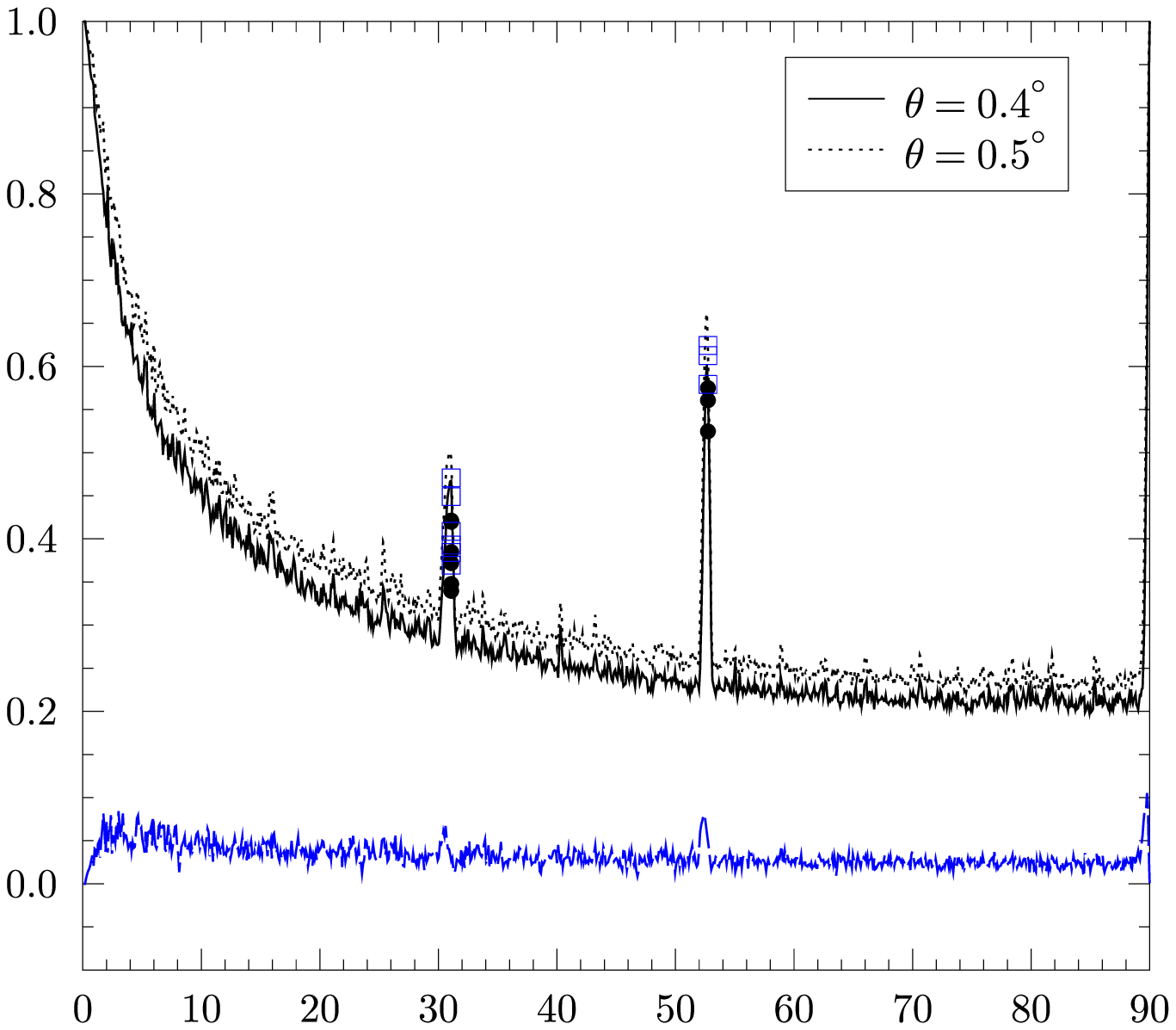}
\put(-36,29){$\alpha$}
\put(-270,195){$S(\alpha)$}
\put(-225,195){(a) V band simulation}
\put(-209,178){$N_{\hbox{\scriptsize side}}=512$}
\end{minipage}
\begin{minipage}{10cm}
\vspace*{-40pt}
\hspace*{-20pt}\includegraphics[width=10.0cm]{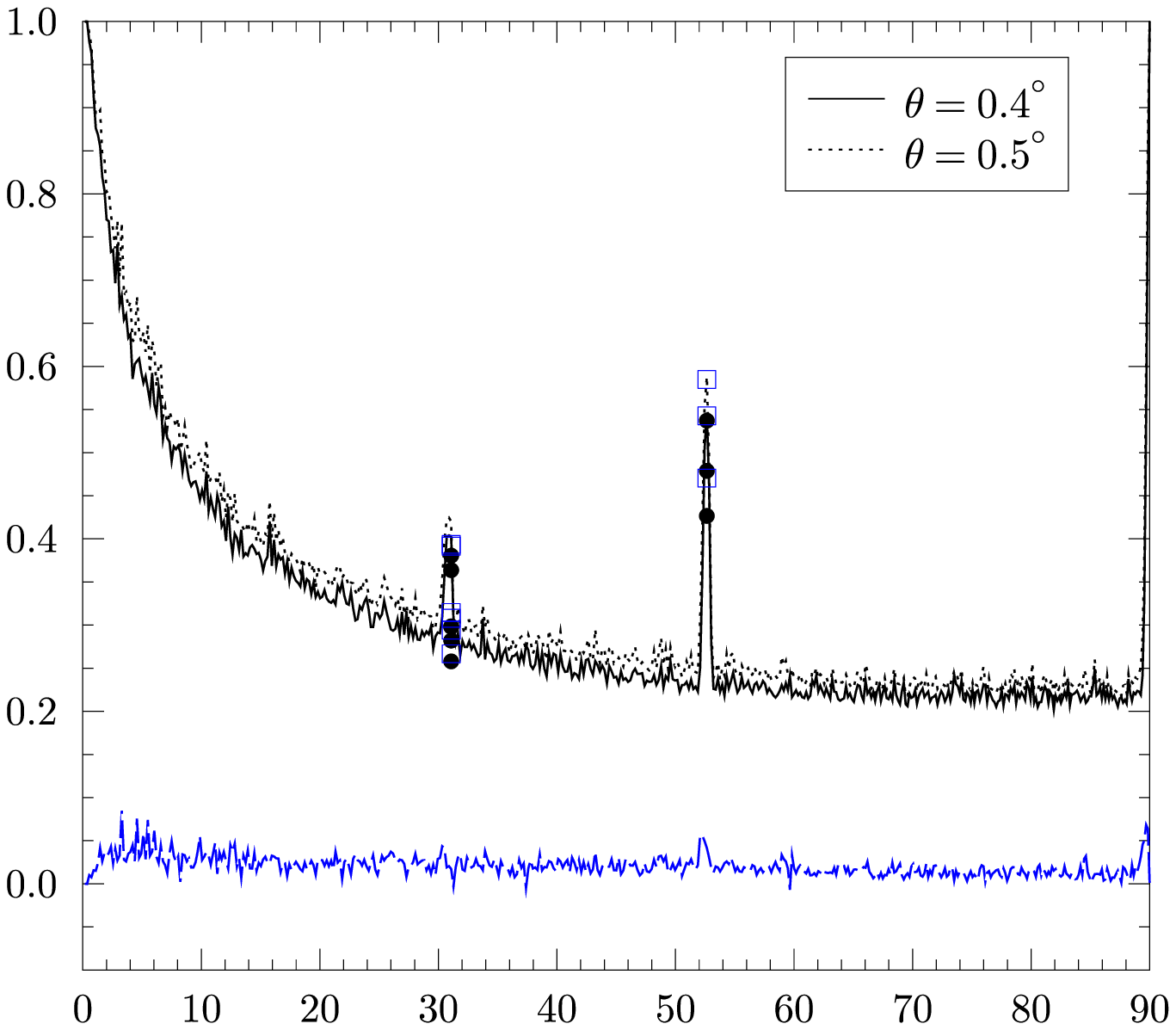}
\put(-36,29){$\alpha$}
\put(-270,195){$S(\alpha)$}
\put(-225,195){(b) V band simulation}
\put(-209,178){$N_{\hbox{\scriptsize side}}=256$}
\end{minipage}
\vspace*{-25pt}
\end{center}
\caption{\label{Fig:CITS_V_W_Noise_Torus}
The CITS correlation $S(\alpha)$ is calculated for the 3-torus simulation
with $L=4.0$ using the beam profile and the detector noise of
the V band.
The CMB map is smoothed using the parameters $\theta=0.4^\circ$ and
$\theta=0.5^\circ$.
In panel (a) $S(\alpha)$ is obtained from the $N_{\hbox{\scriptsize side}}=512$
CMB map, while panel (b) shows the corresponding curves obtained from
the $N_{\hbox{\scriptsize side}}=256$ CMB map.
The difference curve is also shown.
}
\end{figure}

%%%%%%%%%%%%%%%%%%%%%%%%%%%%%%%%%%%%%%%%%%%%%%%%%%%%%%%%%%%%%%%%%%%%%%%%%%%%
%%%%%%%%%%%%%%%%%%%%%%%%%%%%%%%%%%%%%%%%%%%%%%%%%%%%%%%%%%%%%%%%%%%%%%%%%%%%

The importance of the smoothing to reduce the noise
is also revealed by the following analysis.
At first the CMB map with the V-band properties in the
$N_{\hbox{\scriptsize side}}=512$ resolution,
whose CITS correlation is shown in
figure \ref{Fig:CITS_V_W_Torus_Simulation}(a),
is smoothed according to eq.\,(\ref{Eq:Gaussian_beam}).
This smoothing is done in the $N_{\hbox{\scriptsize side}}=512$ resolution
for various values of $\theta$.
The optimum is found in the range between $\theta=0.4^\circ$ and
$\theta=0.5^\circ$.
For smoothing parameters $\theta$ below $\theta=0.4^\circ$
the noise degrades the CITS signal and above $\theta=0.5^\circ$,
the smoothing removes too much CMB information.
The CITS correlation is plotted in figure \ref{Fig:CITS_V_W_Noise_Torus}(a)
for $\theta=0.4^\circ$ and $\theta=0.5^\circ$,
and the improvement compared to figure \ref{Fig:CITS_V_W_Torus_Simulation}(a)
(full curve) is obvious.
In addition, the difference curve is plotted which shows that both smoothing
parameters lead to nearly indistinguishable results.
In panel (b) of figure \ref{Fig:CITS_V_W_Noise_Torus} the analogous
curves are shown for the CMB map which is downgraded to the
$N_{\hbox{\scriptsize side}}=256$ resolution.
Since the computer time is reduced for the smaller
$N_{\hbox{\scriptsize side}}=256$ resolution,
we will use in the following this value of $N_{\hbox{\scriptsize side}}$.

%%%%%%%%%%%%%%%%%%%%%%%%%%%%%%%%%%%%%%%%%%%%%%%%%%%%%%%%%%%%%%%%%%%%%%%%%%%%
%%%%%%%%%%%%%%%%%%%%%%%%%%%%%%%%%%%%%%%%%%%%%%%%%%%%%%%%%%%%%%%%%%%%%%%%%%%%

\begin{figure}
\begin{center}
\begin{minipage}{10cm}
\vspace*{-20pt}
\hspace*{-20pt}\includegraphics[width=10.0cm]{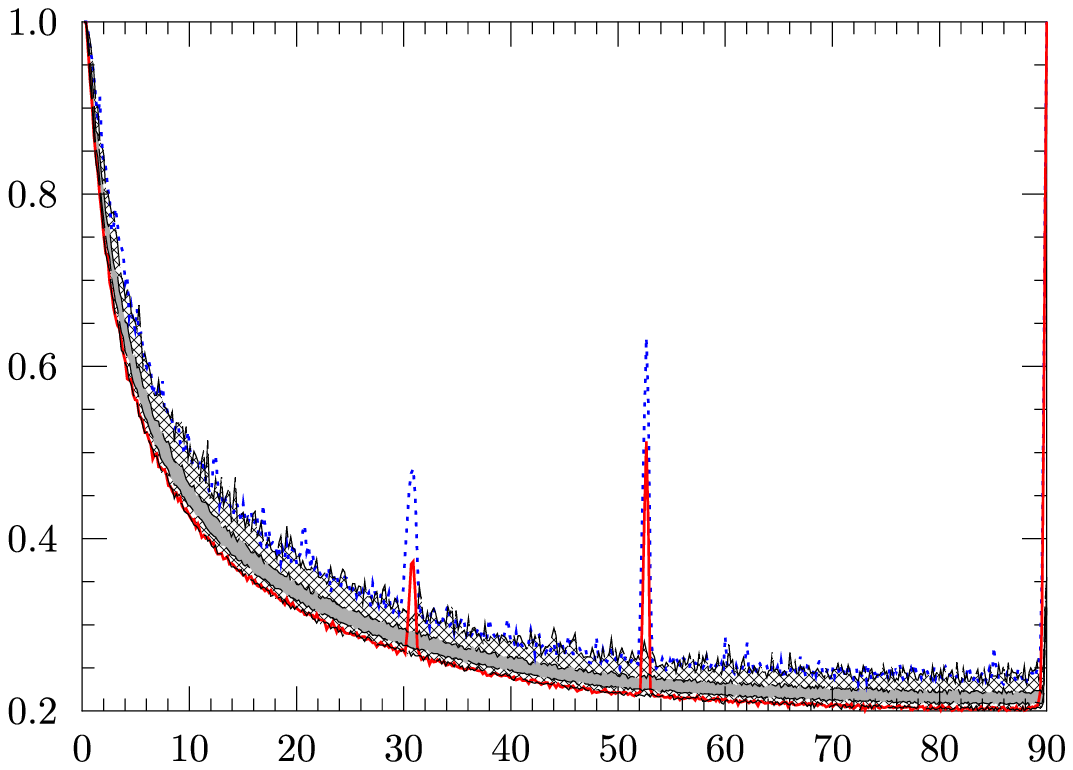}
\put(-35,21){$\alpha$}
\put(-270,153){$S(\alpha)$}
\put(-225,155){(a) 100 V band simulations}
\put(-209,142){$\theta=0.4^\circ$}
\end{minipage}
\begin{minipage}{10cm}
\vspace*{-25pt}
\hspace*{-20pt}\includegraphics[width=10.0cm]{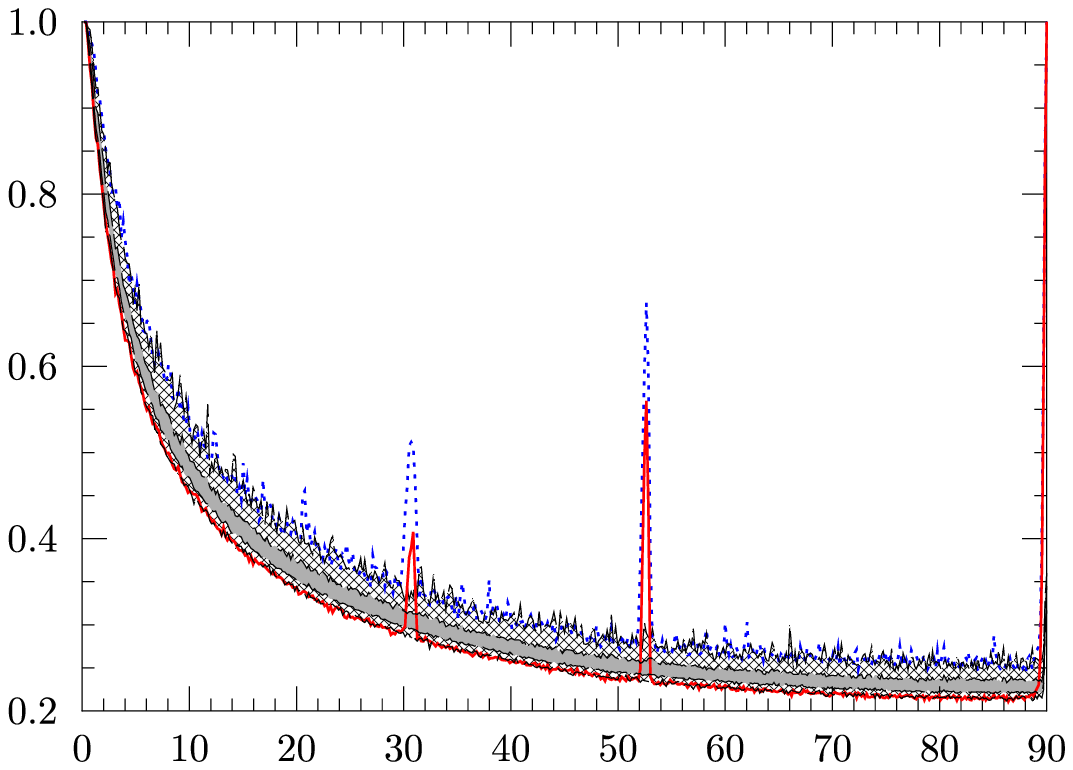}
\put(-35,21){$\alpha$}
\put(-270,153){$S(\alpha)$}
\put(-225,155){(b) 100 V band simulations}
\put(-209,142){$\theta=0.5^\circ$}
\end{minipage}
\vspace*{-25pt}
\end{center}
\caption{\label{Fig:CITS_V_W_Noise_Torus_Statistik}
The CITS correlation $S(\alpha)$ is calculated for 100 different
3-torus simulations with $L=4.0$ using the beam profile and
the detector noise of the V band.
The CMB maps are smoothed using the parameters $\theta=0.4^\circ$ (panel (a))
and $\theta=0.5^\circ$ (panel (b)) and
downgraded to $N_{\hbox{\scriptsize side}}=256$.
The blue dotted curve shows the maximum of $S(\alpha)$ taken over all
100 simulations.
The red full curve shows the corresponding minimum and represents
thus the worst case occurring in the 100 simulations.
The shaded band encompasses all values of $S_{\hbox{\scriptsize bg}}(\alpha)$
occurring in the 100 simulations,
while the grey band contains 66 percent of the $S_{\hbox{\scriptsize bg}}(\alpha)$
values around their median.
This reveals the variation of $S(\alpha)$ with respect to the
random initial conditions of the CMB simulations.
}
\end{figure}

%%%%%%%%%%%%%%%%%%%%%%%%%%%%%%%%%%%%%%%%%%%%%%%%%%%%%%%%%%%%%%%%%%%%%%%%%%%%
%%%%%%%%%%%%%%%%%%%%%%%%%%%%%%%%%%%%%%%%%%%%%%%%%%%%%%%%%%%%%%%%%%%%%%%%%%%%

Up to now the analysis is based on a single CMB simulation
for the 3-torus topology either for $L=1.5$ or for $L=4.0$.
However, the CMB simulations require Gaussian random initial conditions
leading to a statistical ensemble of CMB simulations belonging
to a given topology of a fixed side length $L$.
In order to address the question how this variability affects
the CITS signature, an ensemble of 100 CMB simulations for the 3-torus with
$L=4.0$ is generated.
This leads to 100 CMB maps where the beam profile $b_{V1}(l)$ and
the detector noise of the V band is taken into account as described above.
The results for the two smoothing parameters $\theta=0.4^\circ$
and $\theta=0.5^\circ$ are presented
in figure \ref{Fig:CITS_V_W_Noise_Torus_Statistik}.
The dotted curve shows the largest CITS signature
that occurs in the 100 simulations,
while the full curve shows the worst case
that is the minimum over the 100 CITS correlations $S(\alpha)$.
Individual matched circle pairs can possess even lower values,
but note that $S(\alpha)$ takes the maximum of the 6 pairs at
$\alpha=31.1^\circ$ and of the 3 pairs at $\alpha=52.7^\circ$.
A detailed analysis reveals that the smoothing parameter $\theta=0.5^\circ$
yields marginally better peaks than the choice $\theta=0.4^\circ$.

At radii $\alpha$ where no matched circle pair is nearby,
the correlations $S_{ij}(\alpha,\beta)$,
see eq.\,(\ref{Eq:cits_m_factor}),
show (positive) correlations and (negative) anti-correlations
with the same amplitude, so that the minimum of the anticorrelations,
defined by
\begin{equation}
\label{Eq:cits_avrg}
S_{\hbox{\scriptsize bg}}(\alpha) \; = \;
\max_{i,j,\beta}\,\big( -S_{ij}(\alpha,\beta) \big)
\hspace{10pt} ,
\end{equation}
is a measure of the background even at radii $\alpha$
where matched circle pairs do occur.
Thus it quantifies the detection threshold.
Figure \ref{Fig:CITS_V_W_Noise_Torus_Statistik} shows the
distribution of $S_{\hbox{\scriptsize bg}}(\alpha)$ obtained from the
100 simulations as the shaded band.
This distribution is asymmetric with a tail towards larger values, and
figure \ref{Fig:CITS_V_W_Noise_Torus_Statistik} also shows as the grey
band the band which contains 66 percent of the values of
$S_{\hbox{\scriptsize bg}}(\alpha)$ around their median.

One observes in figure \ref{Fig:CITS_V_W_Noise_Torus_Statistik}
that the CITS signature of the six matched circle pairs at $\alpha=31.1^\circ$
is only slightly above this background in the worst case simulation.
Since the measurements contain residual foregrounds not taken
into account by our simulations, which further deteriorate the signal,
it seems that even the detection of matched circle pairs around
$\alpha=30^\circ$ cannot be assured.
For larger ring radii $\alpha$, see the signature at $\alpha=52.7^\circ$
in figure \ref{Fig:CITS_V_W_Noise_Torus_Statistik},
the situation is far better, and such matched circle pairs should be detected.

%%%%%%%%%%%%%%%%%%%%%%%%%%%%%%%%%%%%%%%%%%%%%%%%%%%%%%
\section{CITS signal for the W and V bands using masks}
%%%%%%%%%%%%%%%%%%%%%%%%%%%%%%%%%%%%%%%%%%%%%%%%%%%%%%
\label{CITS_signal_with_masks}

The noise is not the only problem for the CITS signature,
since foreground sources dominate the CMB in some parts of the sky
so that the CMB signal cannot be reliably reconstructed.
The way out found in the literature consists of masking
the problematic sky regions in the CMB map and replacing the
temperature values by those of the ILC map.
Here, we do not use the pixels which are too severely contaminated
by foregrounds and compute the CITS correlation only with
those pixels which are lying outside the KQ75 and KQ85 masks
provided by the WMAP team \citep{Bennett_et_al_2012}.
Thereby arises the problem that the values of the CITS correlation
cannot be compared for circles having the same radius $\alpha$,
since the correlations are computed from a different number of pixels
depending on the orientation of the circles with respect to the mask.
In a given HEALPix resolution $N_{\hbox{\scriptsize side}}$,
a circle of radius $\alpha$ has the ring index $n_{\hbox{\scriptsize ring}}$
in the RING scheme
\citep{Gorski_Hivon_Banday_Wandelt_Hansen_Reinecke_Bartelmann_2005}
of the HEALPix map,
i.\,e.\ $n_{\hbox{\scriptsize ring}}=n_{\hbox{\scriptsize ring}}(\alpha)$.
Define the number $n_{\hbox{\scriptsize eff}}$ as the number of pixels
that are used for the computation of the CITS correlation,
that is $n_{\hbox{\scriptsize eff}}\leq
n_{\hbox{\scriptsize eff}}^{\hbox{\scriptsize max}}=
4\min(n_{\hbox{\scriptsize ring}},N_{\hbox{\scriptsize side}})$ for
$n_{\hbox{\scriptsize ring}}\leq 2N_{\hbox{\scriptsize side}}$
with equality for the case that the mask does not affect any pixel
on the ring.
The correlation $S(\alpha)$ is then analysed on the two-dimensional grid
$n_{\hbox{\scriptsize eff}}$ versus $n_{\hbox{\scriptsize ring}}$.

According to the results of the preceding sections,
the foreground reduced V- and W-band maps \citep{Bennett_et_al_2012}
of the final 9 year data are expanded in the resolution
$N_{\hbox{\scriptsize side}}=512$ with respect to $a_{lm}$ and
then are smoothed according to eq.\,(\ref{Eq:Gaussian_beam}).
This yields the CMB maps which are analysed in the following
subjected to the KQ75 or KQ85 9 year mask.
The analysis of \cite{Bielewicz_Banday_2011} refers
to the seven year WMAP data subjected to the KQ85 7 year mask.
They take only those circle pairs into account for which less than half
the length of each circle is masked.
Our analysis differentiates much finer the statistical significance
of the masked parts by separating the correlations $S(\alpha)$
with respect to $n_{\hbox{\scriptsize eff}}$.
Thus, only correlations based on a similar number of structures
along a circle are compared because of the equal area pixelization.

%%%%%%%%%%%%%%%%%%%%%%%%%%%%%%%%%%%%%%%%%%%%%%%%%%%%%%%%%%%%%%%%%%%%%%%%%%%%
%%%%%%%%%%%%%%%%%%%%%%%%%%%%%%%%%%%%%%%%%%%%%%%%%%%%%%%%%%%%%%%%%%%%%%%%%%%%

\begin{figure}
\begin{center}
\begin{minipage}{10cm}
\vspace*{-15pt}
\hspace*{-20pt}\includegraphics[width=10.0cm]{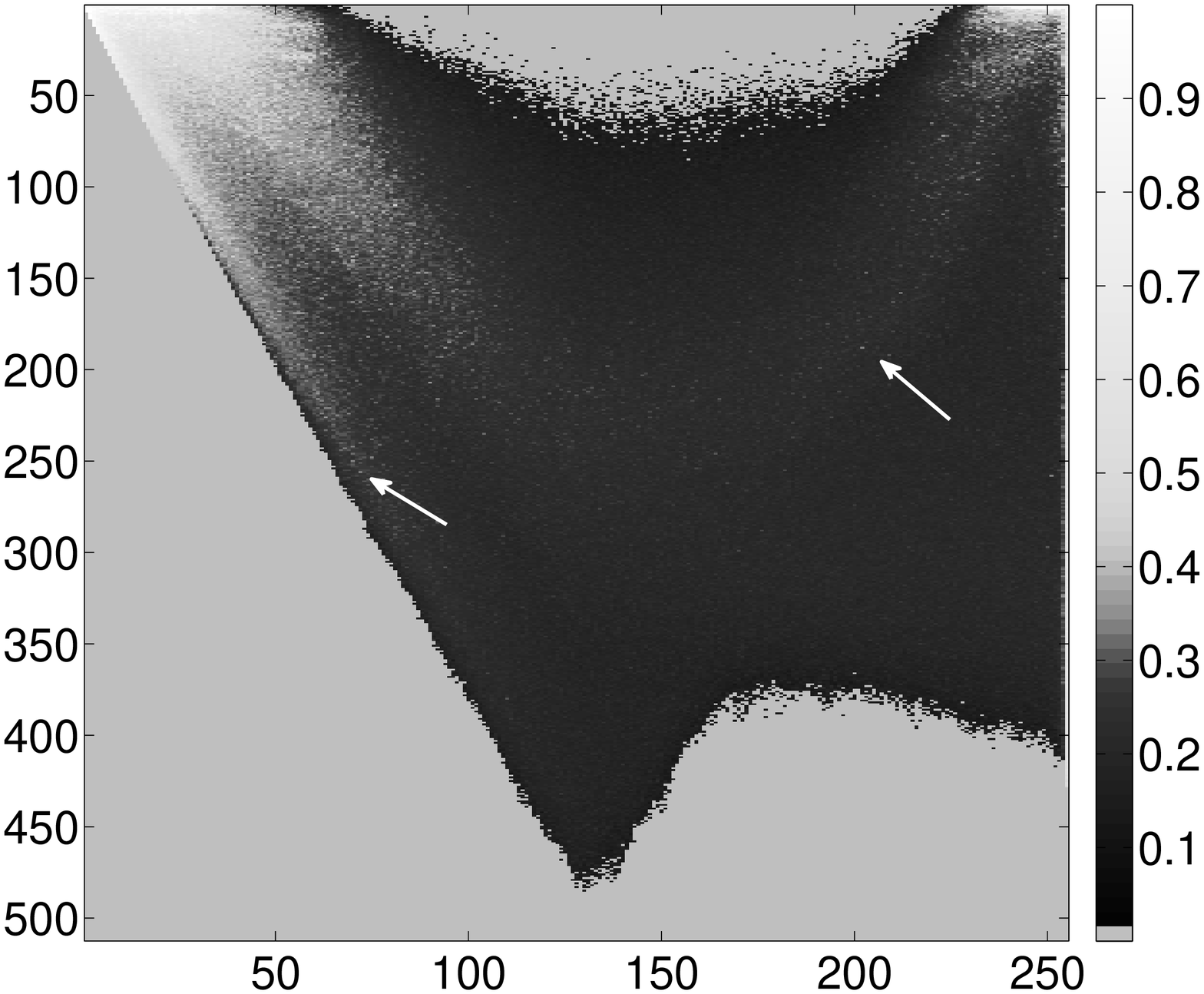}
\put(-80,17){$n_{\hbox{\scriptsize ring}}$}
\put(-270,190){$n_{\hbox{\scriptsize eff}}$}
\put(-240,50){(a) V band WMAP}
\put(-225,35){$N_{\hbox{\scriptsize side}}=128$}
\end{minipage}
\begin{minipage}{10cm}
\vspace*{-10pt}
\hspace*{-20pt}\includegraphics[width=10.0cm]{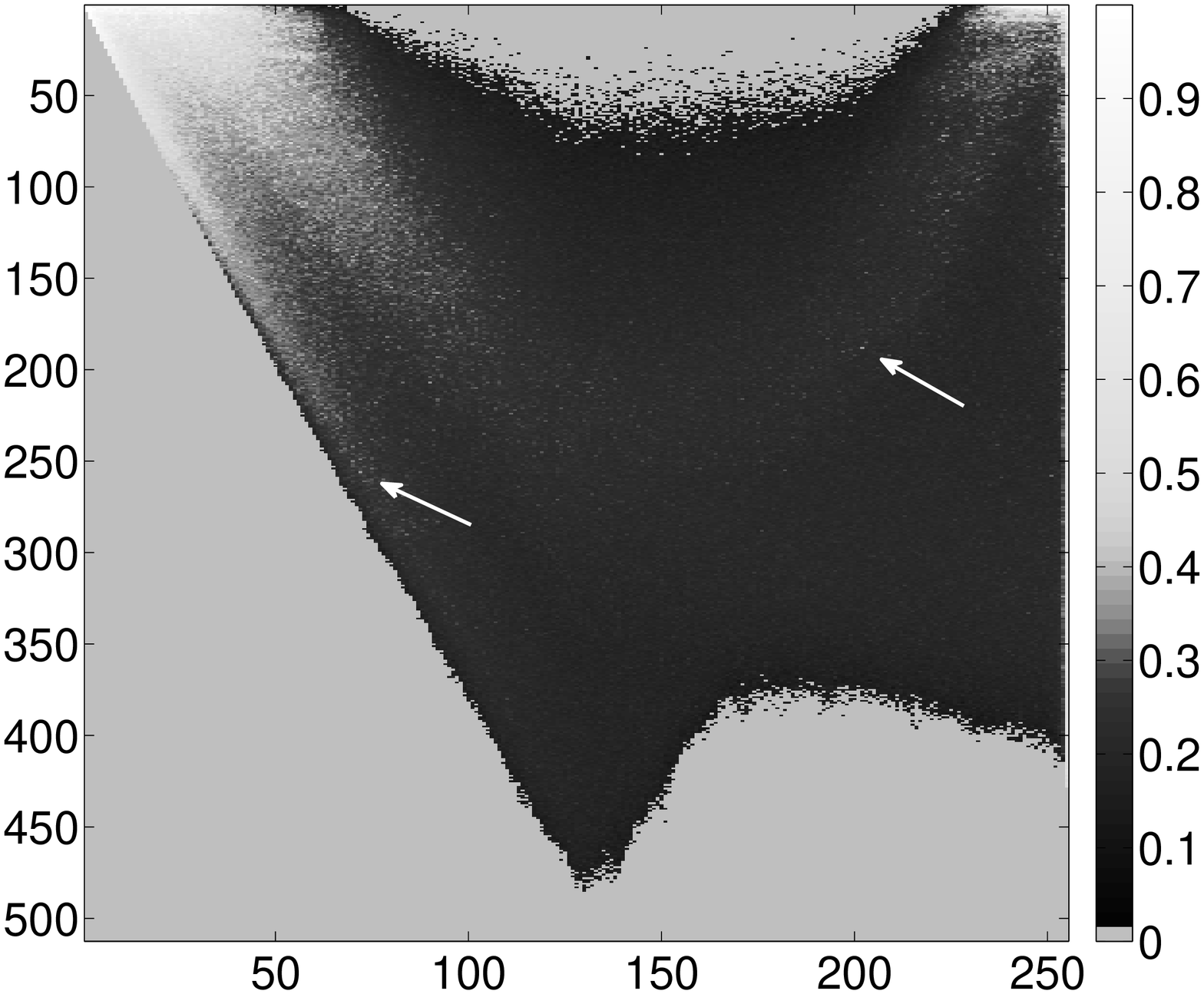}
\put(-80,17){$n_{\hbox{\scriptsize ring}}$}
\put(-270,190){$n_{\hbox{\scriptsize eff}}$}
\put(-240,50){(b) W band WMAP}
\put(-225,35){$N_{\hbox{\scriptsize side}}=128$}
\end{minipage}
\vspace*{-25pt}
\end{center}
\caption{\label{Fig:CITS_V_W_WMAP}
The CITS correlation $S$ is computed for the V and W bands of the
WMAP data using the resolution $N_{\hbox{\scriptsize side}}=128$ as
described in the text.
The maps are smoothed to $0.5^\circ$ and subjected to the KQ75 mask.
The value of $S$ encoded by brightness is shown as a function of
$n_{\hbox{\scriptsize ring}}$ and $n_{\hbox{\scriptsize eff}}$.
Two interesting values are marked by arrows.
}
\end{figure}

%%%%%%%%%%%%%%%%%%%%%%%%%%%%%%%%%%%%%%%%%%%%%%%%%%%%%%%%%%%%%%%%%%%%%%%%%%%%
%%%%%%%%%%%%%%%%%%%%%%%%%%%%%%%%%%%%%%%%%%%%%%%%%%%%%%%%%%%%%%%%%%%%%%%%%%%%

%%%%%%%%%%%%%%%%%%%%%%%%%%%%%%%%%%%%%%%%%%%%%%%%%%%%%%%%%%%%%%%%%%%%%%%%%%%%
%%%%%%%%%%%%%%%%%%%%%%%%%%%%%%%%%%%%%%%%%%%%%%%%%%%%%%%%%%%%%%%%%%%%%%%%%%%%

\begin{figure}
\begin{center}
\begin{minipage}{10cm}
\vspace*{-20pt}
\hspace*{-20pt}\includegraphics[width=10.0cm]{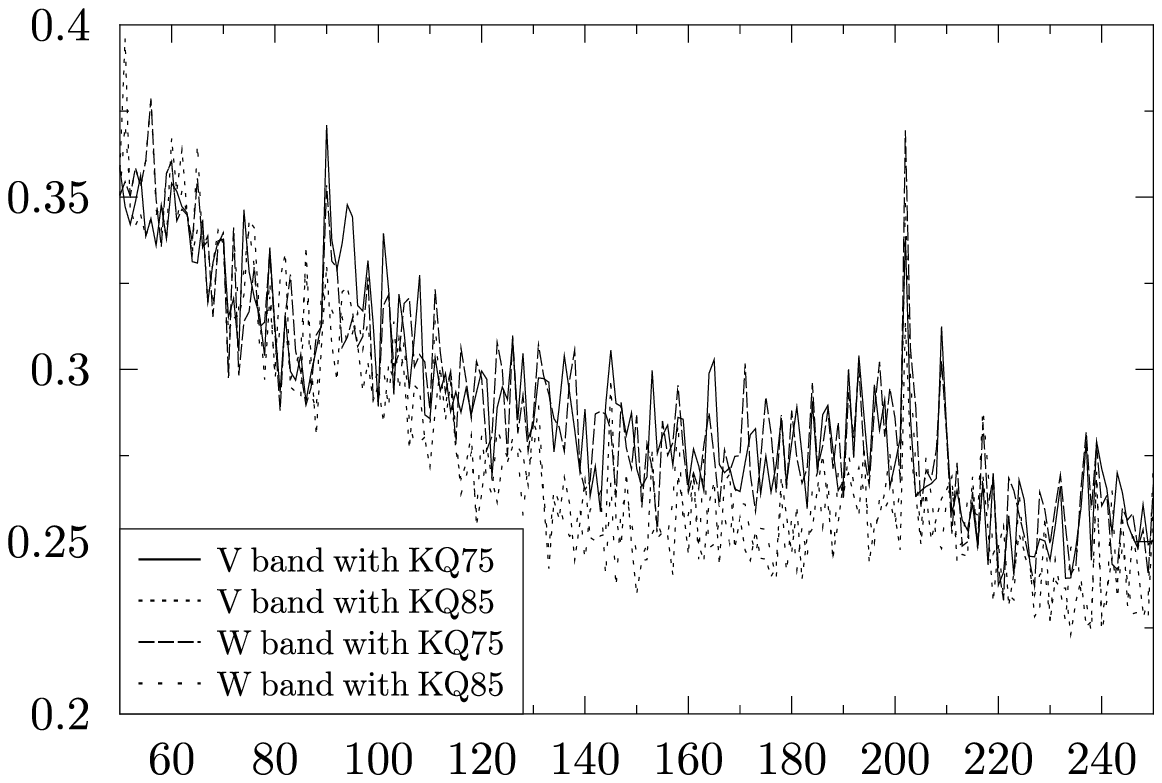}
\put(-42,18){$n_{\hbox{\scriptsize ring}}$}
\put(-260,142){$S$}
\put(-220,147){(a) $n_{\hbox{\scriptsize eff}}=175\dots 250$}
%\put(-190,130){$\theta=0.5^\circ$}
\end{minipage}
\begin{minipage}{10cm}
\vspace*{-25pt}
\hspace*{-20pt}\includegraphics[width=10.0cm]{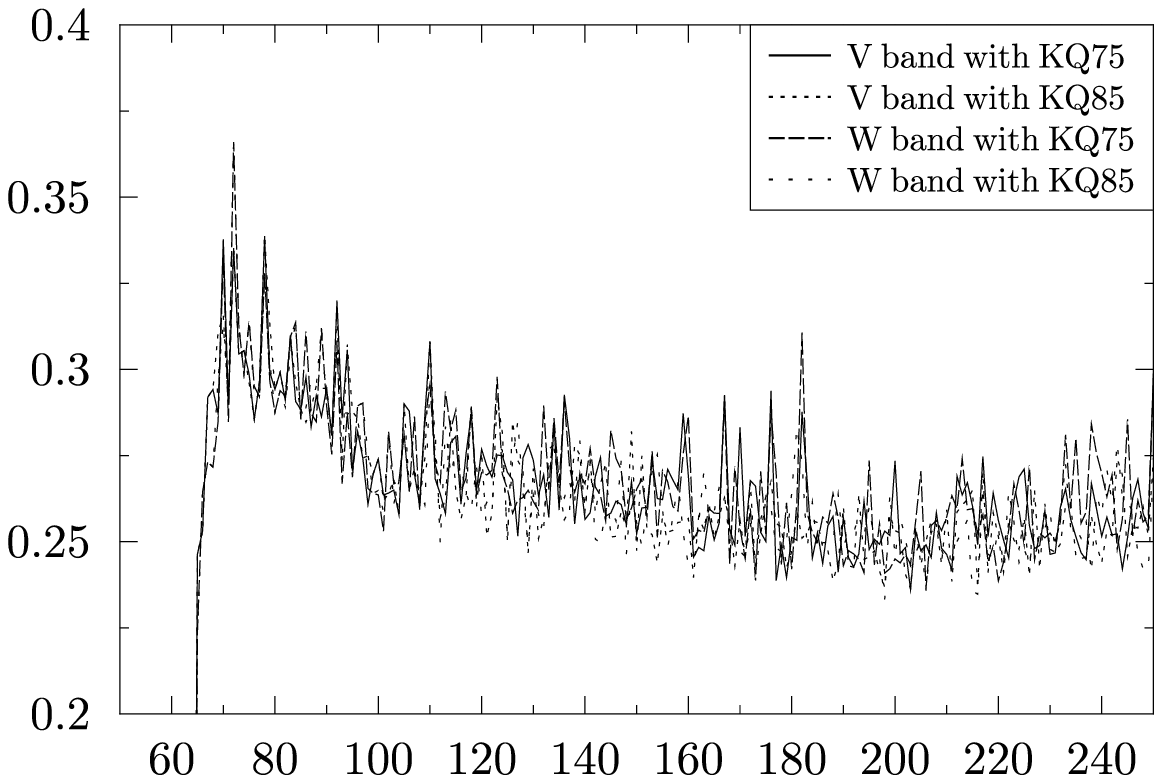}
\put(-42,18){$n_{\hbox{\scriptsize ring}}$}
\put(-260,142){$S$}
\put(-220,147){(b) $n_{\hbox{\scriptsize eff}}=250\dots 500$}
%\put(-214,142){$\theta=0.5^\circ$}
\end{minipage}
\vspace*{-25pt}
\end{center}
\caption{\label{Fig:CITS_V_W_WMAP_neff}
The CITS correlation $S(\alpha)$ shown in figure \ref{Fig:CITS_V_W_WMAP}
is shown as a one-dimensional plot by taking the maximum of $S$ over
the stripe $n_{\hbox{\scriptsize eff}}\in[175,250]$ in panel (a) and over
the stripe $n_{\hbox{\scriptsize eff}}\in[250,500]$ in panel (b).
The peak at $n_{\hbox{\scriptsize ring}}=202$ visible in panel (a) is marked
by the right arrow in figure \ref{Fig:CITS_V_W_WMAP}.
The left arrow corresponds to the peak at $n_{\hbox{\scriptsize ring}}=72$
seen in panel (b).
}
\end{figure}

%%%%%%%%%%%%%%%%%%%%%%%%%%%%%%%%%%%%%%%%%%%%%%%%%%%%%%%%%%%%%%%%%%%%%%%%%%%%
%%%%%%%%%%%%%%%%%%%%%%%%%%%%%%%%%%%%%%%%%%%%%%%%%%%%%%%%%%%%%%%%%%%%%%%%%%%%

We analyse the CMB maps with the resolution $N_{\hbox{\scriptsize side}}=256$
for the smoothing parameters $\theta=0.4^\circ$ and $\theta=0.5^\circ$
by computing the CITS correlation $S$ as a function of
$n_{\hbox{\scriptsize ring}}$ and $n_{\hbox{\scriptsize eff}}$.
We do not find a pronounced peak in the data which could be interpreted as
a signal for a matched circle pair.

Although our tests with the topological simulations discussed in
the previous sections suggest otherwise,
we nevertheless searched in the lower resolution $N_{\hbox{\scriptsize side}}=128$
for peaks.
This differs from the search in the $N_{\hbox{\scriptsize side}}=256$ maps
by putting the average of four $N_{\hbox{\scriptsize side}}=256$ pixels
into one $N_{\hbox{\scriptsize side}}=128$ pixel.
By this additional averaging
the detector noise is suppressed more than necessary according to
our simulations.
However, a possible residual foreground signal on small scales
would also be smoothed out
thereby reducing its contribution to the $m$-weighted circle signature
defined in eq.\,(\ref{Eq:cits_m_factor}).
The figures \ref{Fig:CITS_V_W_WMAP} and \ref{Fig:CITS_V_W_WMAP_neff}
present the results for the CITS search in the $N_{\hbox{\scriptsize side}}=128$
maps.
In figure \ref{Fig:CITS_V_W_WMAP} the value of the CITS correlation
is encoded by brightness.
As already discussed, the horizontal axis $n_{\hbox{\scriptsize ring}}$
corresponds to the radius $\alpha$ of the circle in the
$N_{\hbox{\scriptsize side}}=128$ resolution,
while the vertical axis over $n_{\hbox{\scriptsize eff}}$ separates
the CITS values with respect to the number of pixels
on which its calculation is based.
Towards smaller values of $n_{\hbox{\scriptsize eff}}$, the CITS value
and correspondingly the brightness increases, in general,
because a lower number of pixels enhance the probability
of a large correlation by chance.
A pronounced peak occurs at $n_{\hbox{\scriptsize ring}}=202$
corresponding to $\alpha=73.7^\circ$ and $n_{\hbox{\scriptsize eff}}=188$.
Its position is marked by an arrow in figure \ref{Fig:CITS_V_W_WMAP}.
A further peak at the position $n_{\hbox{\scriptsize ring}}=72$
corresponding to a smaller ring with $\alpha=26.6^\circ$ and
$n_{\hbox{\scriptsize eff}}=256$ is also marked.
Although CITS values belonging to different values of $n_{\hbox{\scriptsize eff}}$
cannot be strictly compared due to their different significance,
we nevertheless combine two stripes with $n_{\hbox{\scriptsize eff}}\in[175,250]$
and $n_{\hbox{\scriptsize eff}}\in[250,500]$ by taking the maximum
in the given $n_{\hbox{\scriptsize eff}}$ interval for a fixed value of
$n_{\hbox{\scriptsize ring}}$.
The results are plotted in figure \ref{Fig:CITS_V_W_WMAP_neff} for both
stripes for the V- and W-band maps subjected to the KQ75 and KQ85 masks.
Panel (a) reveals the pronounced peak at $n_{\hbox{\scriptsize ring}}=202$.
The signal is visible in all 4 cases shown in the figure.
A further peak in panel (a) at $n_{\hbox{\scriptsize ring}}=90$ emerges
only by using the KQ75 mask and is absent using the KQ85 mask.
The decrease in the CITS correlation is thus caused by pixels
close to the boundary of the KQ75 mask.
This case demonstrates that a few additional pixels can destroy a CITS signal.
The peak at $n_{\hbox{\scriptsize ring}}=72$ shown in panel (b) is more clearly
seen in the W band data but is independent of the mask.

%%%%%%%%%%%%%%%%%%%%%%%%%%%%%%%%%%%%%%%%%%%%%%%%%%%%%%%%%%%%%%%%%%%%%%%%%%%%
%%%%%%%%%%%%%%%%%%%%%%%%%%%%%%%%%%%%%%%%%%%%%%%%%%%%%%%%%%%%%%%%%%%%%%%%%%%%

\begin{figure}
\begin{center}
\begin{minipage}{10cm}
\vspace*{-10pt}
\hspace*{0pt}\includegraphics[width=8.0cm]{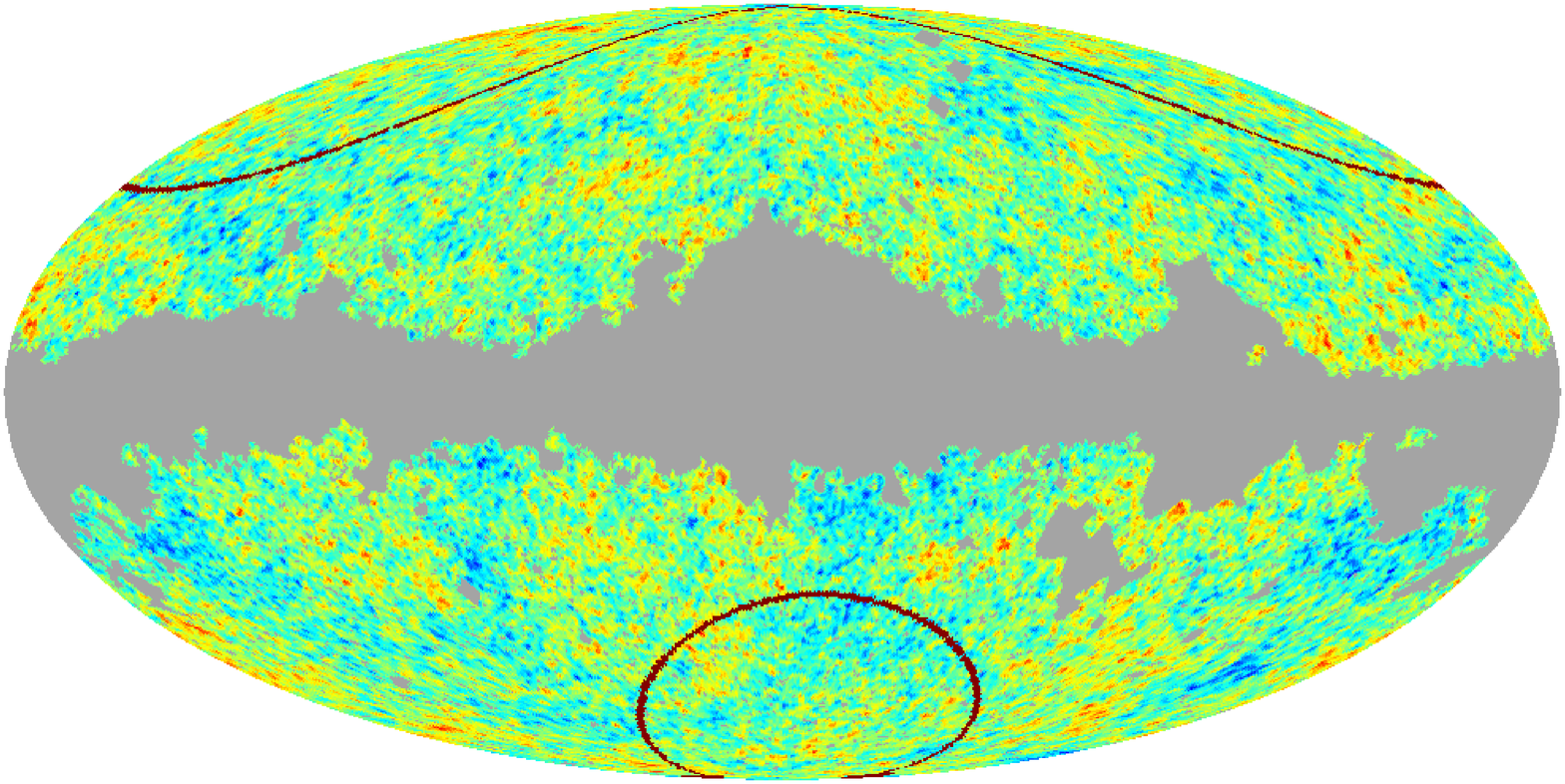}
\put(-230,120){(a) $n_{\hbox{\scriptsize ring}}=72$}
\end{minipage}
\begin{minipage}{10cm}
\vspace*{-10pt}
\hspace*{0pt}\includegraphics[width=8.0cm]{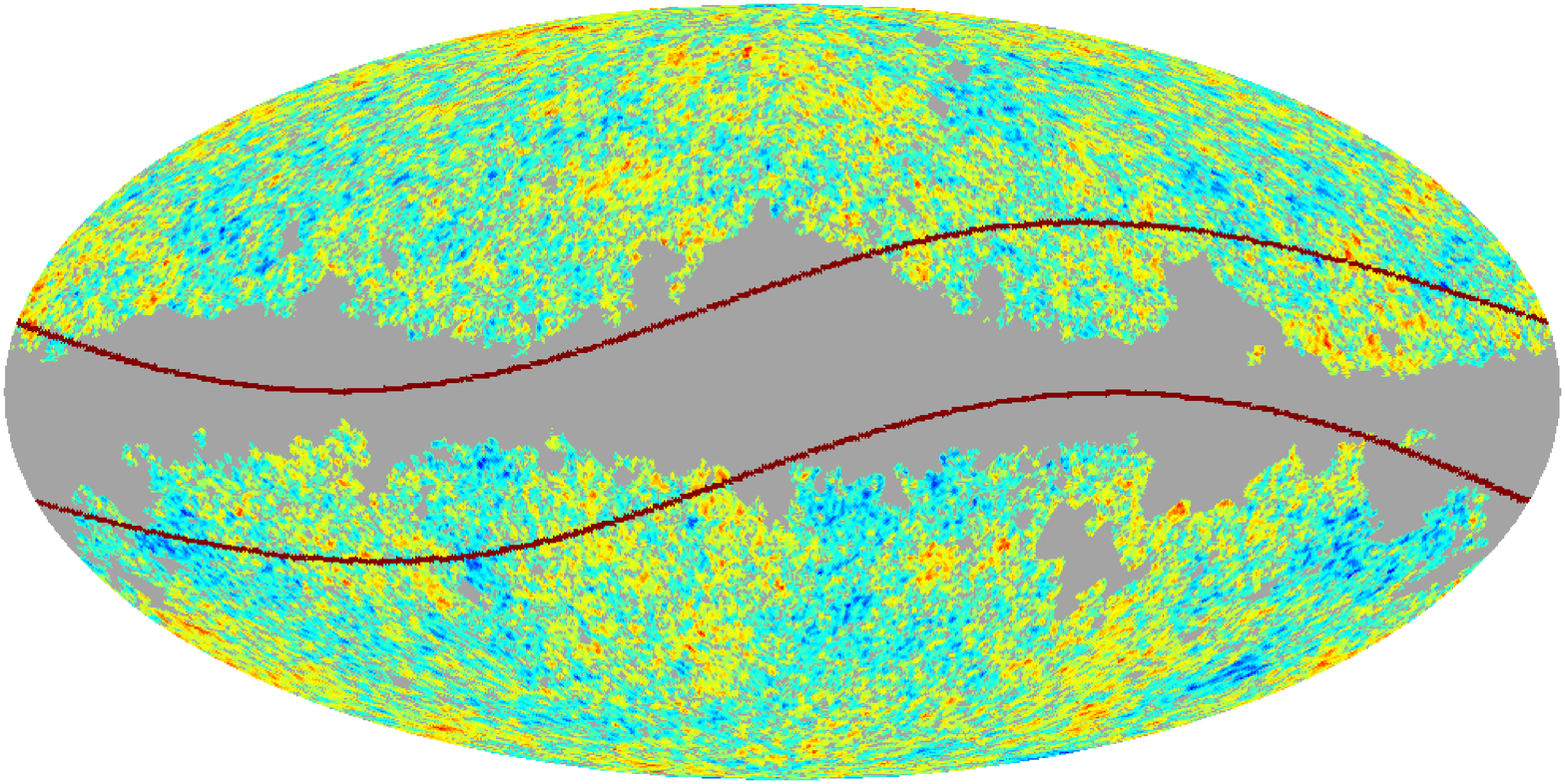}
\put(-230,120){(b) $n_{\hbox{\scriptsize ring}}=202$}
\end{minipage}
\vspace*{-10pt}
\end{center}
\caption{\label{Fig:circle_in_sky_map}
The W band temperature fluctuations subjected to the KQ75 mask are
shown together with the position of the circle pair with
$n_{\hbox{\scriptsize ring}}=72$ (radius $\alpha=26.6^\circ$) in panel (a) and 
with $n_{\hbox{\scriptsize ring}}=202$ (radius $\alpha=73.7^\circ$) in panel (b).
}
\end{figure}

%%%%%%%%%%%%%%%%%%%%%%%%%%%%%%%%%%%%%%%%%%%%%%%%%%%%%%%%%%%%%%%%%%%%%%%%%%%%
%%%%%%%%%%%%%%%%%%%%%%%%%%%%%%%%%%%%%%%%%%%%%%%%%%%%%%%%%%%%%%%%%%%%%%%%%%%%

It is instructive to discuss the position of the circle pair on
the CMB map.
The large circle pair with $n_{\hbox{\scriptsize ring}}=202$ lies in the
neighbourhood of the galactic plane
as shown in figure \ref{Fig:circle_in_sky_map}(b).
This is the reason why large parts of this circle pair are excluded by
the mask.
In addition, this location bears the risk that many pixels used in the
computation of the CITS correlation contain significant amounts of
foreground contributions.
The situation is more favourable in the case of the smaller circle pair with
$n_{\hbox{\scriptsize ring}}=72$ shown in figure \ref{Fig:circle_in_sky_map}(a).
This pair is far away from the galactic plane and most pixels are
taken into account.
At $n_{\hbox{\scriptsize ring}}=72$ one has
$n_{\hbox{\scriptsize eff}}^{\hbox{\scriptsize max}}=288$
which is close to $n_{\hbox{\scriptsize eff}}=256$ found for this circle pair.
On the other hand, the simulations of the preceding sections demonstrate
that a circle pair with such a small radius is at the detection threshold.

%%%%%%%%%%%%%%%%%%%%%%%%%%%%%%%%%%%%%%%%%%%%%%%%%%%%%%%%%%%%%%%%%%%%%%%%%%%%
%%%%%%%%%%%%%%%%%%%%%%%%%%%%%%%%%%%%%%%%%%%%%%%%%%%%%%%%%%%%%%%%%%%%%%%%%%%%

\begin{figure}
\begin{center}
\begin{minipage}{10cm}
\vspace*{-20pt}
\hspace*{-20pt}\includegraphics[width=10.0cm]{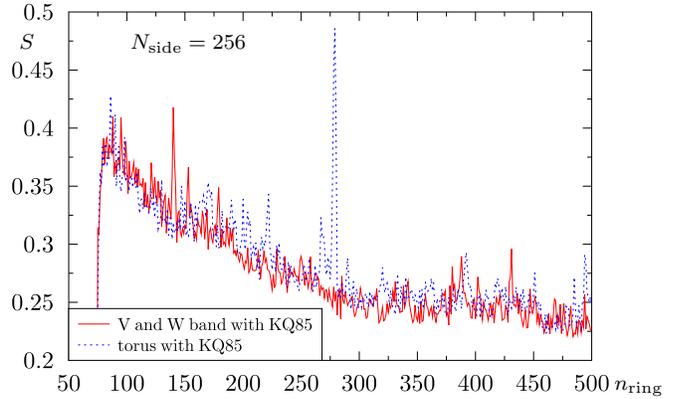}
\put(-35,18){$n_{\hbox{\scriptsize ring}}$}
\put(-260,147){$S$}
\put(-218,147){$N_{\hbox{\scriptsize side}}=256$}
\end{minipage}
\vspace*{-25pt}
\end{center}
\caption{\label{Fig:CITS_VW_WMAP_Torus_neff}
The full curve shows the CITS correlation $S$ obtained from the addition
of the V- and W-band CMB maps,
while the dotted curve belongs to a torus simulation having a
side length $L=4.0$.
The maximum of $S$ over $n_{\hbox{\scriptsize eff}}\geq 300$ is plotted.
The resolution $N_{\hbox{\scriptsize side}}=256$ together with the KQ85 mask
and a smoothing with $\theta=0.5^\circ$ are used.
}
\end{figure}

%%%%%%%%%%%%%%%%%%%%%%%%%%%%%%%%%%%%%%%%%%%%%%%%%%%%%%%%%%%%%%%%%%%%%%%%%%%%
%%%%%%%%%%%%%%%%%%%%%%%%%%%%%%%%%%%%%%%%%%%%%%%%%%%%%%%%%%%%%%%%%%%%%%%%%%%%

The CITS signatures in the resolution $N_{\hbox{\scriptsize side}}=128$
at $n_{\hbox{\scriptsize ring}}=72$ and $n_{\hbox{\scriptsize ring}}=202$ 
are probably generated by chance.
This is suggested by a comparison with the results
in the resolution $N_{\hbox{\scriptsize side}}=256$
where these circle pairs reveal no unusual high correlation values.
Furthermore, we add the V and W band maps after correcting for their
different beam profiles. 
After smoothing the summation map with $\theta=0.5^\circ$,
it is downgraded to $N_{\hbox{\scriptsize side}}=256$.
The maximum of the correlation $S$ with $n_{\hbox{\scriptsize eff}}\geq 300$
is plotted in figure \ref{Fig:CITS_VW_WMAP_Torus_neff} as the full curve,
where the KQ85 mask is used.
The values of $n_{\hbox{\scriptsize ring}}$ for $N_{\hbox{\scriptsize side}}=256$
are twice that of $N_{\hbox{\scriptsize side}}=128$ for the same radii $\alpha$.
So peaks around $n_{\hbox{\scriptsize ring}}=144$ and
$n_{\hbox{\scriptsize ring}}=404$ would be expected from the
$N_{\hbox{\scriptsize side}}=128$ results.
Although a peak at $n_{\hbox{\scriptsize ring}}=140$ appears,
it has nothing to do with that in the $N_{\hbox{\scriptsize side}}=128$ resolution,
since its circle pair has a different orientation.
Furthermore, this peak is absent if the KQ75 mask is used.
The figure \ref{Fig:CITS_VW_WMAP_Torus_neff} also shows as a dotted curve
the CITS correlation for a torus model with $L=4.0$ having circles at
$\alpha=31.1^\circ$
($n_{\hbox{\scriptsize ring}}=168$ for $N_{\hbox{\scriptsize side}}=256$)
and at $\alpha=52.7^\circ$
($n_{\hbox{\scriptsize ring}}=279$ for $N_{\hbox{\scriptsize side}}=256$).
The large circles at $n_{\hbox{\scriptsize ring}}=279$ are clearly revealed,
while none of the six small circle pairs at $n_{\hbox{\scriptsize ring}}=168$
leads to enhanced correlation values.
The fact that the circle pairs at $\alpha=31.1^\circ$ are not detected,
suggests that the peak at $n_{\hbox{\scriptsize ring}}=140$
corresponding to $\alpha=25.8^\circ$ is produced by chance and possible
contributions from foreground sources.
Furthermore, this fact also shows that matched circle pairs should have
radii larger than $\alpha=30^\circ$ in order to be detectable
with WMAP data.

%%%%%%%%%%%%%%%%%%%%%%%%%%%%%%%%%%%%%%%%%%%%%%%%%%%%%%%%%%%%%%%%%%%%%%%%%%%%
%%%%%%%%%%%%%%%%%%%%%%%%%%%%%%%%%%%%%%%%%%%%%%%%%%%%%%%%%%%%%%%%%%%%%%%%%%%%

\begin{figure}
\begin{center}
\begin{minipage}{10cm}
\vspace*{-20pt}
\hspace*{-20pt}\includegraphics[width=10.0cm]{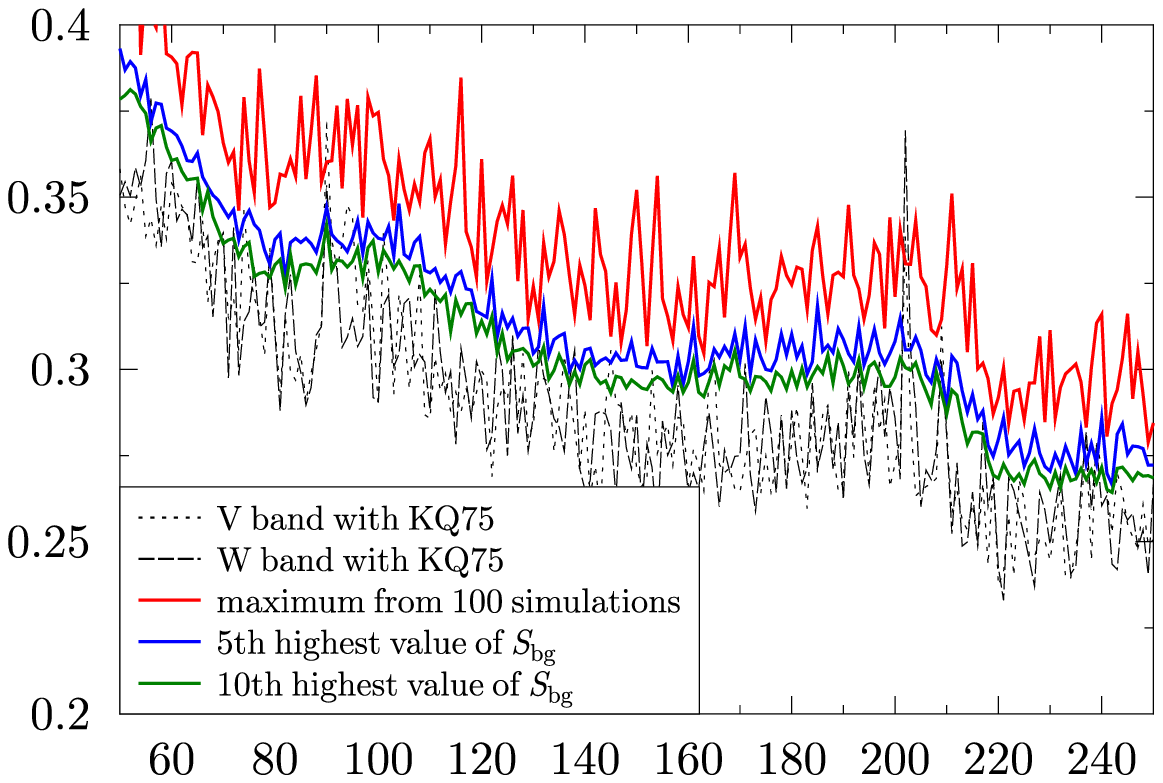}
\put(-42,18){$n_{\hbox{\scriptsize ring}}$}
\put(-260,142){$S$}
\put(-175,148){(a) $n_{\hbox{\scriptsize eff}}=175\dots 250$}
%\put(-190,130){$\theta=0.5^\circ$}
\end{minipage}
\begin{minipage}{10cm}
\vspace*{-25pt}
\hspace*{-20pt}\includegraphics[width=10.0cm]{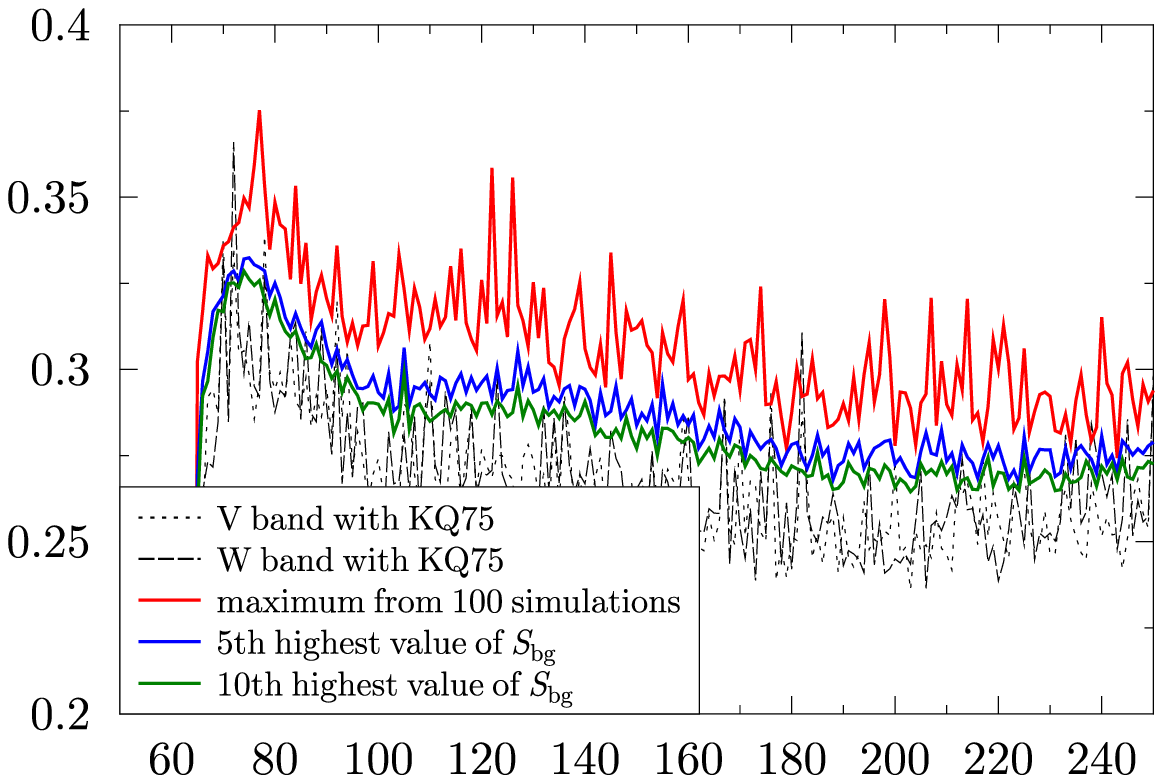}
\put(-42,18){$n_{\hbox{\scriptsize ring}}$}
\put(-260,142){$S$}
\put(-175,148){(b) $n_{\hbox{\scriptsize eff}}=250\dots 500$}
%\put(-214,142){$\theta=0.5^\circ$}
\end{minipage}
\vspace*{-22pt}
\end{center}
\caption{\label{Fig:CITS_V_W_WMAP_neff_kq75}
The CITS correlation for the V- and W-band data subjected to the KQ75 mask,
as also shown in figure \ref{Fig:CITS_V_W_WMAP_neff}, is compared with
$S_{\hbox{\scriptsize bg}}$ obtained from 100 simulations.
The resolution $N_{\hbox{\scriptsize side}}=128$ is used.
The topmost full curve shows the maximum of $S_{\hbox{\scriptsize bg}}$
that occurs among these 100 simulations.
The next lower full curve shows the 5th highest value of
$S_{\hbox{\scriptsize bg}}$ among the 100 simulations,
while the lowest full curve displays the 10th highest value.
The stripe $n_{\hbox{\scriptsize eff}}\in[175,250]$ is plotted in panel (a) and
the stripe $n_{\hbox{\scriptsize eff}}\in[250,500]$ in panel (b).
}
\end{figure}

%%%%%%%%%%%%%%%%%%%%%%%%%%%%%%%%%%%%%%%%%%%%%%%%%%%%%%%%%%%%%%%%%%%%%%%%%%%%
%%%%%%%%%%%%%%%%%%%%%%%%%%%%%%%%%%%%%%%%%%%%%%%%%%%%%%%%%%%%%%%%%%%%%%%%%%%%

%%%%%%%%%%%%%%%%%%%%%%%%%%%%%%%%%%%%%%%%%%%%%%%%%%%%%%%%%%%%%%%%%%%%%%%%%%%%
%%%%%%%%%%%%%%%%%%%%%%%%%%%%%%%%%%%%%%%%%%%%%%%%%%%%%%%%%%%%%%%%%%%%%%%%%%%%

\begin{figure}
\begin{center}
\begin{minipage}{10cm}
\vspace*{-20pt}
\hspace*{-20pt}\includegraphics[width=10.0cm]{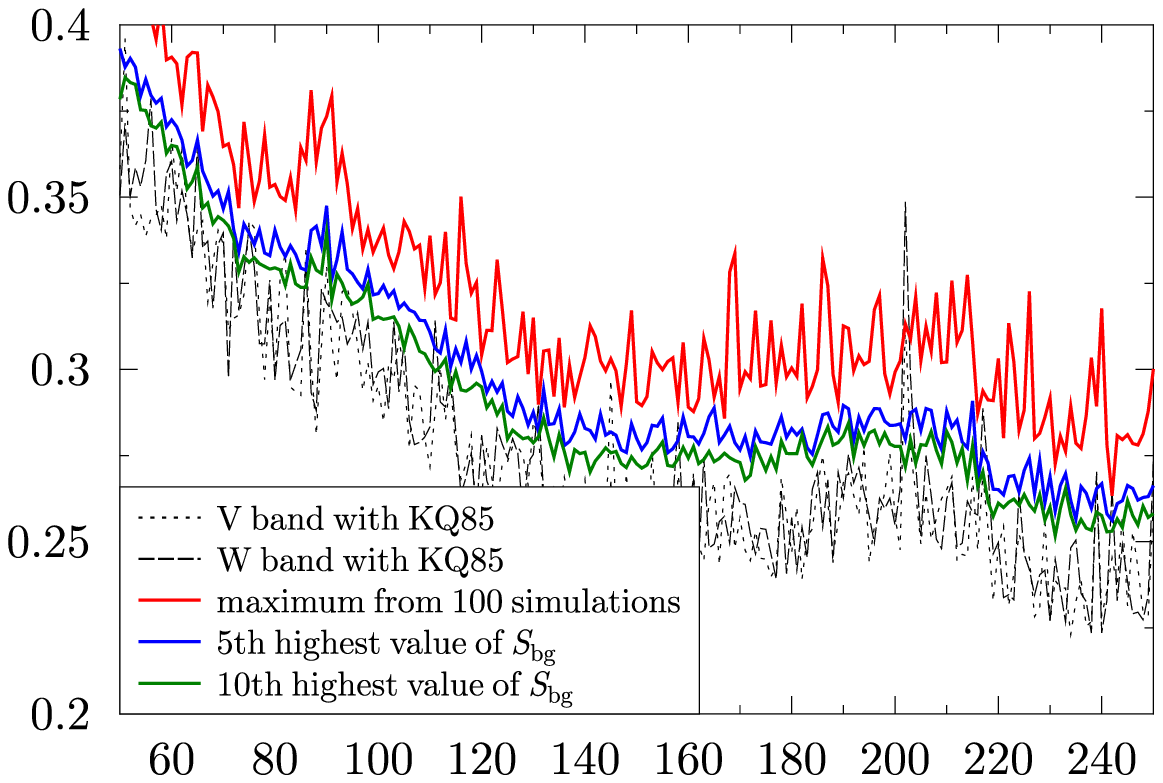}
\put(-42,18){$n_{\hbox{\scriptsize ring}}$}
\put(-260,142){$S$}
\put(-175,148){(a) $n_{\hbox{\scriptsize eff}}=175\dots 250$}
%\put(-190,130){$\theta=0.5^\circ$}
\end{minipage}
\begin{minipage}{10cm}
\vspace*{-25pt}
\hspace*{-20pt}\includegraphics[width=10.0cm]{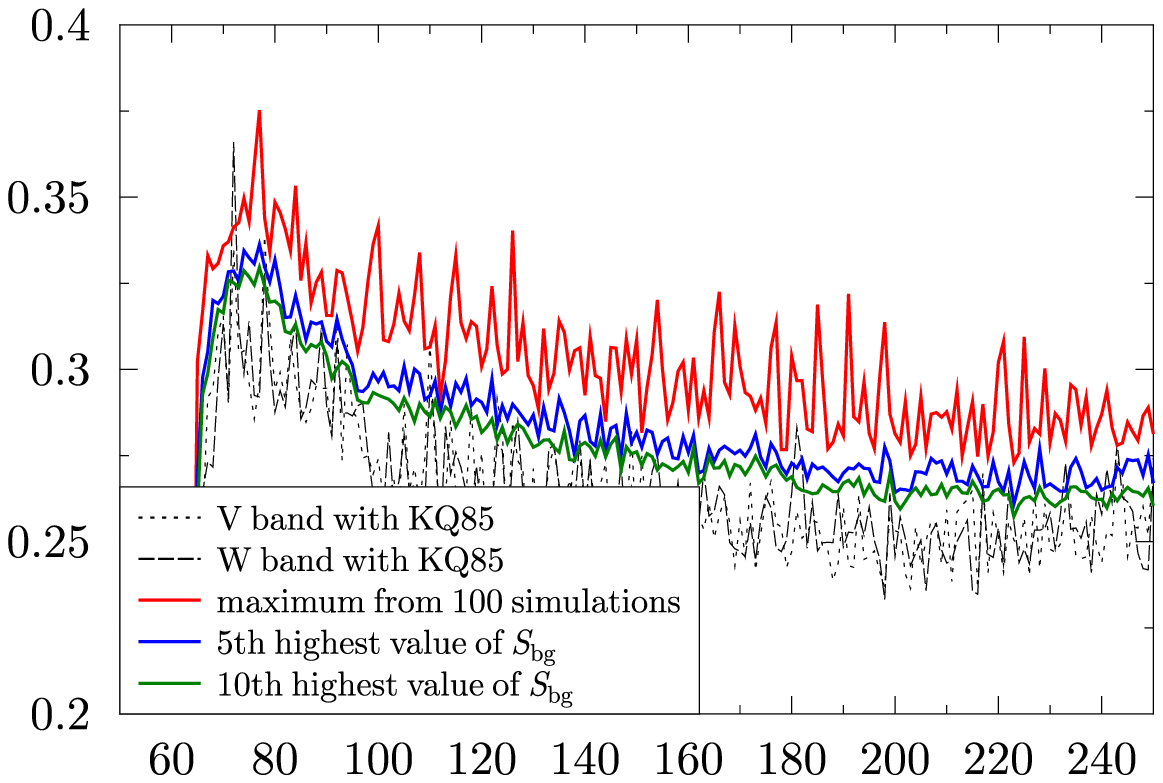}
\put(-42,18){$n_{\hbox{\scriptsize ring}}$}
\put(-260,142){$S$}
\put(-175,148){(b) $n_{\hbox{\scriptsize eff}}=250\dots 500$}
%\put(-214,142){$\theta=0.5^\circ$}
\end{minipage}
\vspace*{-22pt}
\end{center}
\caption{\label{Fig:CITS_V_W_WMAP_neff_kq85}
The analogous curves as in figure \ref{Fig:CITS_V_W_WMAP_neff_kq75}
are plotted which are here computed using the KQ85 mask.
}
\end{figure}

%%%%%%%%%%%%%%%%%%%%%%%%%%%%%%%%%%%%%%%%%%%%%%%%%%%%%%%%%%%%%%%%%%%%%%%%%%%%
%%%%%%%%%%%%%%%%%%%%%%%%%%%%%%%%%%%%%%%%%%%%%%%%%%%%%%%%%%%%%%%%%%%%%%%%%%%%

Although the CITS peaks of the two matched circle candidates cannot be
confirmed in the $N_{\hbox{\scriptsize side}}=256$ resolution,
it is instructive to compare the $N_{\hbox{\scriptsize side}}=128$
CITS signature with the 100 simulations
that were used in section \ref{Noise_CITS_amplitude},
see also figure \ref{Fig:CITS_V_W_Noise_Torus_Statistik},
to estimate the detection threshold in the case without mask.
These 100 3-torus simulations take the beam profile and the detector noise
of the V band into account, and in the following,
only the smoothing $\theta=0.5^\circ$ is considered.
The 100 CMB maps were downgraded to $N_{\hbox{\scriptsize side}}=128$,
and the background amplitude $S_{\hbox{\scriptsize bg}}$,
see eq.\,(\ref{Eq:cits_avrg}), is computed for each of the 100 maps.
The maximum taken from the 100 values of $S_{\hbox{\scriptsize bg}}$ estimates
the detection threshold in the sense
that if a CITS signal higher than this threshold occurs,
its probability would be lower than 1:100 in comparison with the simulations.
In contrast to the detection threshold without mask,
shown in figure \ref{Fig:CITS_V_W_Noise_Torus_Statistik},
the maximum over the 100 $S_{\hbox{\scriptsize bg}}$ curves is no longer
a relatively smooth curve, but instead shows large fluctuations
which are caused by the selective behaviour of the mask.
Therefore, also the 5th and 10th highest values of the 100
$S_{\hbox{\scriptsize bg}}$ values are computed leading to much smoother curves.
The results are presented in figures \ref{Fig:CITS_V_W_WMAP_neff_kq75}
and \ref{Fig:CITS_V_W_WMAP_neff_kq85} for the KQ75 9yr and KQ85 9yr mask,
respectively.
One observes that both matched circle pair candidates,
the one belonging to $n_{\hbox{\scriptsize ring}}=72$ corresponding to
$\alpha=26.6^\circ$,
and the other one belonging to $n_{\hbox{\scriptsize ring}}=202$
with $\alpha=73.7^\circ$, take on values larger than $S_{\hbox{\scriptsize bg}}$
in any of the 100 simulations.
Comparing these two peaks in the CITS correlation with the 5th or the
10th highest $S_{\hbox{\scriptsize bg}}$ value reveals an even higher excess.
Furthermore, the results do not seem to depend on the chosen mask,
since the KQ85 mask as well as the more restrictive KQ75 mask leads
to similar conclusions.
Note that the 5th or the 10th highest $S_{\hbox{\scriptsize bg}}$ value
provides an upper envelope of the V- and W-band results with only a
few exceptions.
This demonstrates that the height of the two peaks is indeed unusual.
However, since the peaks possess no correspondence in the higher resolved
$N_{\hbox{\scriptsize side}}=256$ CMB maps, they are probably not a genuine
topological signature.
As shown above, our simulations with the characteristics of the V- and W-band
maps reveal a stronger topological signature in the
$N_{\hbox{\scriptsize side}}=256$ maps contrary to the behaviour of the two
CITS candidates.
A loophole in this conclusion might be
that the V- and W-band maps could possess residual foregrounds
on such small scales that their degrading contribution to the CITS signal is
almost eliminated in the low-resolution $N_{\hbox{\scriptsize side}}=128$ CMB maps
but not in $N_{\hbox{\scriptsize side}}=256$ CMB maps.

%%%%%%%%%%%%%%%%%%%%%%%%%%%%%%%%%%%%%%%%%%%%%%%%%%%%%%
\section{Summary and discussion}
%%%%%%%%%%%%%%%%%%%%%%%%%%%%%%%%%%%%%%%%%%%%%%%%%%%%%%

The circles-in-the-sky (CITS) signature provides a test for a
possible non-trivial topology, i.\,e.\ for a multiply connected space,
of our Universe.
The detection threshold of a matched circle pair depends on
the radius $\alpha$ of the circle.
In order to determine the detection threshold, accurate CMB maps are simulated
for the 3-torus topology, and their CITS correlations are computed for 
various circumstances.

The CITS correlation is introduced in eqs.\,(\ref{Eq:cits_m_factor}) and
(\ref{Eq:cits}) in section \ref{Individual_CITS_amplitude}.
This topological detection measure is applied to a simulated 3-torus CMB map
for a restricted search for back-to-back circle pairs.
It turns out that for radii $\alpha \gtrsim 25^\circ$,
the CITS signal is strong enough for a detection of matched circle pairs.
Furthermore, it is found that the highest correlation values are not
found on the matched circles, but rather for neighbouring circles
having slightly smaller radii.
The smaller the radii of the matched circle pairs,
the more pronounced is this effect.

The experiments provide CMB maps in which the resolution is determined
by the size of the pixels.
Section \ref{Smoothing_CITS_amplitude} investigates the question
how an additional Gaussian smoothing of the CMB data can enhance the
detection probability for a matched circle pair.
The amplitude of the CITS signal is compared with the detection threshold
for several CMB simulations with different HEALPix resolutions
$N_{\hbox{\scriptsize side}}$ and different Gaussian smoothing parameters
$\theta$, see eq.\,(\ref{Eq:Gaussian}).
This analysis leads to the conclusion that for $N_{\hbox{\scriptsize side}}=256$
a smoothing parameter of $\theta=0.4^\circ$ is preferable
if the pixels contain the clean CMB signal.
But for $N_{\hbox{\scriptsize side}}=512$ one should prefer a smaller
smoothing parameter $\theta$ in order to detect also matched circle pairs
with smaller radii.
Experimentally, such clean sky maps are not available because of
detector noise and residual foregrounds.

Before these problems are addressed,
we turn in section \ref{search_grid} to the question how sharp the peaks are
which belong to genuine matched circle pairs.
This determines the resolution of the search grid $N_{\hbox{\scriptsize grid}}$
whose coordinates are used as possible circle centres for
which the CITS correlation is computed.
It is demonstrated that for a smoothing of $\theta=0.4^\circ$,
several matched circle pairs get missed at $N_{\hbox{\scriptsize grid}}=128$
while using $N_{\hbox{\scriptsize side}}=512$ for the CMB map.
Although one can find matched circle pairs at $N_{\hbox{\scriptsize grid}}=128$,
one cannot exclude matched circle pairs if no circle pairs are found at
this search grid resolution.

In sections \ref{Individual_CITS_amplitude} to \ref{search_grid}
only general properties are studied without reference to a specific
experiment.
In section \ref{Noise_CITS_amplitude}, however, CMB maps
for the 3-torus topology are generated which take the noise properties
of the V- and W-band channels of the WMAP satellite into account.
The CITS correlation for a CMB map having the quality
corresponding to the WMAP 9yr observations is computed in the
$N_{\hbox{\scriptsize side}}=512$ resolution.
It turns out that the CITS signal of the matched circle pairs is either absent
or very faint which demonstrates that the noise is too severe in order
to search directly in the original V- and W-band maps for a topological
signature.
This result is further substantiated by smoothing the
$N_{\hbox{\scriptsize side}}=512$ map using the parameters
$\theta=0.4^\circ$ and $\theta=0.5^\circ$.
These additionally smoothed $N_{\hbox{\scriptsize side}}=512$ maps reveal
the CITS peaks at the positions of the matched circle pairs.
Such a smoothing is necessary for a sufficient suppression of noise.
Since the above conclusions are based on a single CMB simulation,
a set of 100 CMB maps is generated for the 3-torus topology with
side length $L=4$.
Their 100 CITS correlations reveal the variability with respect to
the Gaussian random initial conditions which determine how the eigenmodes
are weighted.
As shown in figure \ref{Fig:CITS_V_W_Noise_Torus_Statistik},
the variability of the height of the CITS peaks at circle radii
around $\alpha\simeq 30^\circ$ is so important
that some matched circle pairs are at the  detection limit.
Around $\alpha\simeq 50^\circ$ the variability is of the same order,
but the detection probability is not affected
due to the more pronounced peaks.

The results of sections \ref{Individual_CITS_amplitude} to
\ref{Noise_CITS_amplitude} are obtained from simulations for only
two different side lengths $L$ of the 3-torus topology.
Although the magnitude of the CITS correlation does not directly depend on
the side length $L$,
the detection probability depends on it through the simple fact
that smaller topological sizes lead to more matched circle pairs
which in turn enhances the discovery probability.
The dependence on the cosmological parameters is also modest
because of the small range of variability available for them
due to the current accurate measurements.
Furthermore, the results apply only to back-to-back searches
since the general search leads to a much higher background level
because of the vast increase in possible combinations of circle pairs.

Section \ref{CITS_signal_with_masks} addresses as a further complication
the residual foregrounds which can be so severe that it is advisable
to mask the unsafe regions.
As a consequence, the CITS correlation can only be computed from the unmasked
fraction of the circle leading to a statistical significance depending
on the number $n_{\hbox{\scriptsize eff}}$ of pixels used for the
computation of the CITS correlation.
The search for a CITS signature in the foreground reduced V- and W-band maps
with $N_{\hbox{\scriptsize side}}=256$ and $\theta=0.4^\circ$ or $\theta=0.5^\circ$
subjected to the KQ75 or KQ85 mask yields a negative result.
Although the results obtained from our simulation would not advise to
use the $N_{\hbox{\scriptsize side}}=128$ resolution in a CITS search,
we nevertheless carry out the $N_{\hbox{\scriptsize side}}=128$ search
and find some interesting peaks.
A peak emerges from the background at the large circle radius
$\alpha = 73.7^\circ$ as seen in figures \ref{Fig:CITS_V_W_WMAP},
\ref{Fig:CITS_V_W_WMAP_neff}(a), \ref{Fig:CITS_V_W_WMAP_neff_kq75}(a)
and \ref{Fig:CITS_V_W_WMAP_neff_kq85}(a).
The corresponding circle pair lies, however,
as seen in figure \ref{Fig:circle_in_sky_map}(b),
very close to the galactic plane.
Since one expects the largest residual foregrounds close to the galactic plane,
the suspicion arises that the signal might be spuriously generated
by residual foregrounds.
However, if one accepts such an argument, the reverse argument
that these very residual foregrounds can also destroy a true CITS signal,
has to be accepted in the same way.
Furthermore, since so many pixels are masked in this circle pair,
the high CITS correlation could also be produced by chance.
But the simulations presented in figures \ref{Fig:CITS_V_W_WMAP_neff_kq75}(a)
and \ref{Fig:CITS_V_W_WMAP_neff_kq85}(a) demonstrate that the amplitude
of the peak is indeed unusual.
A second interesting peak in the CITS correlation belongs to
a circle radius $\alpha = 26.6^\circ$,
which is also probably generated by chance as our simulations show.
On the other hand, this circle pair is far from the galactic plane,
and few pixels are masked and the residual foregrounds are less important.
As demonstrated in section \ref{CITS_signal_with_masks},
the six circle pairs at $\alpha=31.1^\circ$ in the $L=4.0$ torus topology
are not detected by using the quality of the WMAP data,
which shows that matched circle pairs have to be larger
than at least $\alpha=30^\circ$ in order to be detectable.
We are thus eagerly awaiting new more accurate data in order to see,
whether the CITS peaks will withstand or prove to be a spurious result.
Even if these CITS signatures will not endure, a multiply connected space
can possess circle pairs that are not back-to-back and
these are not considered in this paper.
Examples for topologies with such generic matched circle pairs
are provided by the half-turn space $E_2$ \citep{Aurich_Lustig_2010b}
and by spherical double-action manifolds
\citep{Aurich_Lustig_2012c,Aurich_Lustig_2012d}.
These spaces would require an extended search for generic matched circle pairs.

%%%%%%%%%%%%%%%%%%%%%%%%%%%%%%%%%%%%%%%%%%%%%%%%%%%%%%%%%%%%%%%%%%%%%%%%%%%%
%%%%%%%%%%%%%%%%%%%%%%%%%%%%%%%%%%%%%%%%%%%%%%%%%%%%%%%%%%%%%%%%%%%%%%%%%%%%

\section*{Acknowledgements}

HEALPix [healpix.jpl.nasa.gov]
\citep{Gorski_Hivon_Banday_Wandelt_Hansen_Reinecke_Bartelmann_2005}
and the WMAP data from the LAMBDA website (lambda.gsfc.nasa.gov)
were used in this work.

%%%%%%%%%%%%%%%%%%%%%%%%%%%%%%%%%%%%%%%%%%%%%%%%%%%%%%%%%%%%%%%%%%%%%%%%%%%%
%%%%%%%%%%%%%%%%%%%%%%%%%%%%%%%%%%%%%%%%%%%%%%%%%%%%%%%%%%%%%%%%%%%%%%%%%%%%

% \newblock {\em arXiv:}1201.6490 [astro-ph.CO].
% \newblock {\em arXiv:}1103.3505 [astro-ph.CO].

\bibliography{../bib_astro}
%\bibliography{../bib_astro_mnras}
%\bibliographystyle{mn2e}
\bibliographystyle{hapalike}

\label{lastpage}

\end{document}